\pdfoutput=1

\documentclass[11pt]{article}

\usepackage[final]{acl}

\usepackage{times}
\usepackage{latexsym}

\usepackage[T1]{fontenc}

\usepackage[utf8]{inputenc}

\usepackage{microtype}

\usepackage{inconsolata}

\usepackage{graphicx}

\title{PersonaX: A Recommendation Agent-Oriented User Modeling Framework for Long Behavior Sequence}

\author{
  \textbf{Yunxiao Shi\textsuperscript{1}},
  \textbf{Wujiang Xu\textsuperscript{2}},
  \textbf{Zeqi Zhang\textsuperscript{1}},
\\
  \textbf{Xing Zi\textsuperscript{1}},
  \textbf{Qiang Wu\textsuperscript{1}},
  \textbf{Min Xu\textsuperscript{1}}
\\
  \textsuperscript{1}University of Technology Sydney, 
  \textsuperscript{2}Rutgers University
\\
\small\texttt{{\href{mailto:Yunxiao.Shi@student.uts.edu.au}{Yunxiao.Shi@student.uts.edu.au}, \href{mailto:Min.Xu@uts.edu.au}{Min.Xu@uts.edu.au}
 }}
}

\usepackage{algorithm}
\usepackage{algpseudocode}
\usepackage{amssymb} 
\usepackage{amsmath}
\usepackage{amsthm}
\usepackage{booktabs}
\usepackage{multirow}
\usepackage{algpseudocode}
\usepackage{graphicx}

\usepackage{tcolorbox}
\usepackage{xcolor}
\usepackage{svg}
\usepackage{bm}
\usepackage{afterpage}
\usepackage{mdframed}
\usepackage{caption}

\theoremstyle{plain}

    \newtheorem{lemma}{Lemma}
    \newtheorem{proposition}{Proposition}

\theoremstyle{definition} 
  \newtheorem{definition}{Definition}

\definecolor{background_u}{RGB}{240, 240, 240}
\definecolor{frame_u}{RGB}{200, 200, 200}
\definecolor{define_pink}{rgb}{0.988, 0.541, 0.416} 
\definecolor{define_gray}{rgb}{0.502, 0.502, 0.502}

\newsavebox{\InterviewCase}

\begin{document}
\definecolor{background_u}{HTML}{FEF9F5}
\definecolor{frame_u}{HTML}{98450E}
\definecolor{background_i}{HTML}{F9FBFD}
\definecolor{frame_i}{HTML}{2E75B5}
\definecolor{background_e}{HTML}{FAFAFA}
\definecolor{frame_e}{HTML}{0D0D0D}

\maketitle
\begin{abstract}
User profile embedded in the prompt template of personalized recommendation agents play a crucial role in shaping their decision-making process. High-quality user profiles are essential for aligning agent behavior with real user interests. Typically, these profiles are constructed by leveraging LLMs for user profile modeling (LLM-UM). However, this process faces several challenges: (1) LLMs struggle with long user behaviors due to context length limitations and performance degradation. (2) Existing methods often extract only partial segments from full historical behavior sequence, inevitably discarding diverse user interests embedded in the omitted content, leading to incomplete modeling and suboptimal profiling. (3) User profiling is often tightly coupled with the inference context, requiring online processing, which introduces significant latency overhead. \textbf{In this paper, we propose PersonaX, an agent-agnostic LLM-UM framework to address these challenges. It augments downstream recommendation agents to achieve better recommendation performance and inference efficiency.} PersonaX (a) segments complete historical behaviors into clustered groups, (b) selects multiple sub-behavior sequences (SBS) with a balance of prototypicality and diversity to form a high-quality core set, (c) performs offline multi-persona profiling to capture diverse user interests and generate fine-grained, cached textual personas, and (d) decouples user profiling from online inference, enabling profile retrieval instead of real-time generation. \textbf{Extensive experiments demonstrate its effectiveness: using only 30–50\% of behavioral data (sequence length 480), PersonaX enhances AgentCF by 3–11\% and Agent4Rec by 10–50\%.} As a scalable and model-agnostic LLM-UM solution, PersonaX sets a new benchmark in scalable user modeling. 
The code is available at URL~\footnote{https://github.com/Ancientshi/PersonaX}.
\end{abstract}

\vspace{-4mm}
\section{Introduction}
Recent advances in LLMs \cite{Gpt4tools, Toolllm, a-mem, iagent} have enabled instruction-based agents \cite{iagent}  to excel in autonomous interaction and decision-making \cite{CAMEL,wang2024survey,RCAgent}. By integrating realistic user profiles into prompts, these agents achieve personalization and effectively mimick real user behaviors \cite{better_agent, character_agent}. Personalized recommendation agents—such as AgentCF~\cite{AgentCF}, Agent4Rec~\cite{Agent4Rec}, and RecAgent~\cite{RecAgent}—inherit this potential yet face a challenge: users seldom state their preferences explicitly, leaving them implicit in their historical behavioural traces. Hence, modeling representative descriptive user profiles from implicit feedback becomes crucial for unleashing the full power of personalized recommendation agents.

\begin{figure*}[t]
  \centering
    \includegraphics[width=6.0 in]{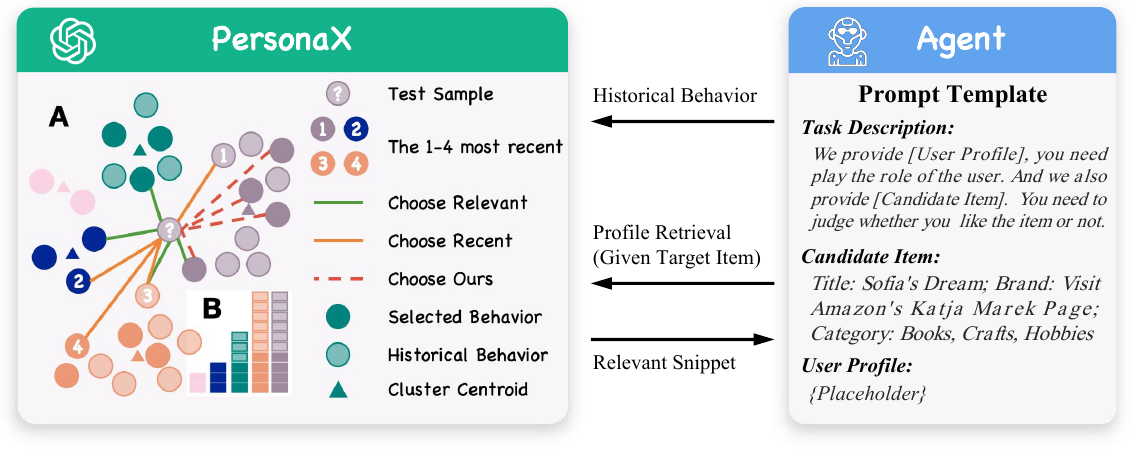}
    \vspace{-3mm}
   \caption{PersonaX as a tool for user modeling given historical behaviors. PersonaX deliver retrieved persona snippet to downstream agent for decision making. In the left part, Persona.A illustrates the behavior distribution, clustering results, and selected/unselected samples. Persona.B presents distribution of the sampling budgets allocation results at a 50\% selection ratio.}
   \vspace{-6mm}
    \label{fig: demo}
\end{figure*}

Recommendation agents typically employ large language models for real-time user modelling (LLM-UM). The profile produced by the LLM is embedded in the prompt and guides the model when generating recommendations for a target item. Existing LLM-UM methods can be grouped into three categories. \textbf{Demonstration} approaches encode the user’s behavior sequence (BS) directly into the prompt as in-context examples, allowing the LLM to generalize from explicit demonstrations \cite{search_based_UM, uncovering, chatgptgoodrecommender}. \textbf{Summarization} techniques  distill extensive interaction histories into concise textual personas that capture core preference signals; this strategy, shown to improve personalization \cite{Summarization}, is adopted by ONCE \cite{ONCE} and Agent4Rec \cite{Agent4Rec}. Moreover, methods like AgentCF \cite{AgentCF} and RecAgent \cite{RecAgent} adopt a \textbf{Reflection} approach, employing reflection mechanisms \cite{lift_up, ExpeL, ACN} on the behavior sequence to iteratively refine user persona.

LLM-UM depends on historical behavioral sequences (BS) that encode rich preference signals. A straightforward strategy is to employ the \textbf{Full data} into the LLM for profiling \cite{Agent4Rec}, but this quickly becomes infeasible as sequence length grows. \textbf{Recent sampling} truncates the user’s behavior sequence to its most recent interactions, prioritizing short-term interests \cite{llmrank, AgentCF}. This strategy depends on temporal information that are only available during online inference. Consequently, both the sampling  and the subsequent LLM-UM process must be performed online. Alternatively, \textbf{Relevance sampling} \cite{lamp, Agent4Rec, personalized_search_memory} selects those past behaviors most pertinent with the target item for capturing long-term preference patterns. Like Recent sampling, it relies on item-specific contextual signals that only become available at inference time—so this strategy also must be executed online.

We identify three principal limitations in the way current LLM-UM methods utilize historical behavioral data: (1) \textbf{Difficulty profiling from long behavior sequences.} Summarization-based approaches are constrained by LLMs’ maximum input length and suffer from the “lost-in-the-middle” phenomenon \cite{LongRAG,Shi2024,ERAGent,improve_by_retrieve,RAG}, whereby critical mid-sequence context is omitted, undermining accurate preference inference. Reflection-based methods, in turn, incur prohibitive computational cost and latency when reasoning over very lengthy sequences. Furthermore, excessive behavioral data introduce noise and redundancy that obscure truly salient signals. (2) \textbf{Sampling inevitably incurs information loss.} By omitting valuable behavioral signals, existing sampling strategies can compromise the quality of the generated user profile. (3) \textbf{Profiling relies on online contextual information and results in inference latency.} Both recent and relevance sampling strategies require real-time contextual inputs (e.g., current timestamp, target item), which mandates modeling user profiles at inference time and incurs decision-making latency overhead.

To address these challenges, we propose \textbf{PersonaX}, a novel LLM-UM framework that performs end-to-end profile from long behavioral sequences in an offline setting, cached for online retrieval by downstream recommendation agents for decision making. Such paradigm significantly reduces online inference latency and enhances overall recommendation performance, their architecture is illustrated in Figure~\ref{fig: demo}. PersonaX segments full historical behaviors into clustered groups and selects sub-behavior sequences (SBS) for each cluster with prototypicality-diversity balanced sampling. Summarization or Reflection is conducted on SBS in an offline manner, generating multiple fine-grained personas that capture diverse user interests. Those cached textual representations are then retrieved by downstream recommendation agents during online inference stage. PersonaX prioritizes example quality over quantity, using only a small fraction of behavioral data (sequence length < 5) to select compact yet informative SBS and avoid selecting irrelevant or noisy samples, thus overcoming issues such as input length constraints and mid-content oversight. Compared to both Recent and Relevance sampling, our method constructs a high-fidelity core-set that preserves the full spectrum of user interests, thereby avoiding information loss. PersonaX assumes responsibility for profile modeling, enabling cached profile retrieval instead of on‐the‐fly generation and thereby eliminating application agents’ online inference latency.

In summary, our contributions are threefold. (1) \textbf{PersonaX Framework}: We introduce PersonaX, an LLM-based user-profile modeling framework oriented for recommendation agents. By decoupling profile generation from online inference, PersonaX eliminates real-time modeling overhead—accelerating inference—and delivers more representative user profiles that substantially boost downstream recommendation performance. (2) \textbf{Data-Efficient Core Behavior Selection}: We introduce a novel strategy for selecting core behaviors via clustering, adaptive allocation of sampling budgets, and a prototypicality-diversity-balanced in-cluster selection mechanism. By using only 30–50\% of the data utilization and ignore other redundant behaviors, our method generates multiple compact sub‐behavior sequences (SBSs), each capturing a distinct facet of user preferences.
(3) \textbf{Extensive Validation}: We evaluate PersonaX with two leading recommendation agents—AgentCF and Agent4Rec— on next-item ranking tasks across three datasets of varying sequence lengths. PersonaX consistently boosts ranking accuracy (3–11\% for AgentCF; 10–50\% for Agent4Rec) and accelerates online inference efficiency.

\section{Preliminary}
\subsection{User Modeling}
Let 
$
\mathcal{S} = \{(I_1,L_1), (I_2,L_2), \dots, (I_n,L_n)\} 
$
denotes a user's historical behavior sequence of length $ n $,
where $ I_i $ represents the $ i $-th interacted item and $ L_i \in \{0,1\} $ indicates the corresponding interaction label ($ 0 $ for dislike and $ 1 $ for like). We define the task of user modeling is to construct a precise and representative user persona $ \mathcal{P}(\mathcal{S}) $ by leveraging the historical behavioral data $ \mathcal{S} $, where $ \mathcal{P}(\cdot) $ is a user modeling method (e.g., Summarization and Reflection). The learned user persona should capture the implicit preference patterns underlying interactions, enabling augmentation for downstream instructional agent recommendation.

\subsection{Sub-Behavior Sequence (SBS) Selection.}
To tackle the challenge of LLM-UM struggling with analyzing long behavior sequence, sampling methods are often employed on the full historical sequence $\mathcal{S}$. These methods aim to extract a Sub-Behavior Sequence (SBS) that retains the most essential information necessary for accurate user profiling while significantly reducing sequence length. Formally, let $ \mathcal{S}^* = \{\hat{I}_{1}, \hat{I}_{2}, \dots, \hat{I}_{k}\} \subseteq \mathcal{S} $ denote the SBS of length $ k $ ($ k \ll n $), where $ \hat{I}_{i} $ represents the $ i $-th selected behavior. The selection ratio, $\frac{k}{n} $, quantifies the compression achieved.

\section{Method}
\subsection{Behavior Clustering}
\label{sec: clustering}
We employ hierarchical clustering to group items based on user interest similarity, treating each cluster as a cohesive analysis unit. A language embedding model $\mathbf{E}(\cdot)$, such as BGE Embedding \cite{BGE_embedding} or EasyRec \cite{EasyRec}, encodes each item $ I_i $ into a dense vector $\mathbf{e}_i$. Let $\mathcal{E} = \{\mathbf{e}_1, \mathbf{e}_2, \dots, \mathbf{e}_n\}$ represent the item embeddings from the user’s interaction history. Pairwise similarity is measured via Euclidean distance: $ d(\mathbf{e}_i, \mathbf{e}_j) = \|\mathbf{e}_i - \mathbf{e}_j\|_2 $, denoted as $ d_{i,j} $.  

Clustering is controlled by a distance threshold $ \tau $, which restricts the maximum intra-cluster distance while preventing merges between clusters with inter-cluster distances below $ \tau $. The resulting clusters $\mathcal{C} = \{c_1, c_2, \dots, c_m\}$ satisfy Intra-cluster constraint: $\forall c \in \mathcal{C}, \forall I_i, I_j \in c, d_{i,j} < \tau$ and Inter-cluster constraint: $\forall c_p, c_q \in \mathcal{C}, c_p \neq c_q, d(c_p, c_q) \geq \tau$.

\subsection{Sampling Budget Allocation}
Given a finite budget $ k $ for sampling historical behaviors, we propose a Sampling Budget Allocation Strategy to distribute this budget across clusters. The algorithm dynamically adjusts allocation based on cluster size distribution, ensuring that smaller clusters are given sufficient attention while preventing larger clusters from dominating the selection process. This promotes a balanced distribution of selected samples, preserving the diversity of sampled behaviors and maintaining a representative coverage of the data \cite{CCS}.

The strategy first sorts clusters by size in ascending order. Each cluster is initially assigned an average allocation $ q $. If a cluster’s size is smaller than $ q $, it receives its exact size, and $ q $ is recalculated based on the remaining quota. Otherwise, the cluster is allocated $ q $. This process repeats iteratively until the entire budget is assigned. Algorithm~\ref{alg:SamplingBudgetAllocation} details the method, and Figure~\ref{fig: demo}.B illustrates an example, where smaller clusters are fully allocated first, and the remaining budget is evenly distributed among larger clusters.

\begin{algorithm}[t]
\footnotesize
\caption{Sampling Budget Allocation} \label{alg:SamplingBudgetAllocation}
\begin{algorithmic}[1]
\State \textbf{Input:} Set of clusters $\mathbf{C} = \{c_1, c_2, \dots, c_m\}$, Cluster size list $\mathbf{s} = \{s_1, s_2, \dots, s_m\}$ where $s_i = |c_i|$, Total sampling budget $k$
\State \textbf{Output:} Allocation list $\mathbf{A} = \{a_1, a_2, \dots, a_m\}$

\Function{AllocateBudget}{$\mathbf{C}, \mathbf{s}, k$}
    \State Sort $\mathbf{s}$ in ascending order and obtain sorted indices $\mathbf{I}$
    \State Initialize allocation $\mathbf{A} \gets [0, 0, \dots, 0]$ 
    \State Remaining budget $B \gets k$

    \For{each cluster $i$ in sorted order}
        \State $r \gets$ number of remaining clusters
        \State $q \gets B \mathbin{//} r$ \Comment{Average allocation per remaining cluster}
        \State $a_i \gets \min(s_i, q)$ \Comment{Allocate min of cluster size or $q$}
        \State $B \gets B - a_i$ \Comment{Update remaining budget}
    \EndFor

    \While{$B > 0$}  \Comment{Distribute any remaining budget}
        \For{each cluster $i$ \textbf{in sorted order} \textbf{if} $a_i < s_i$}
            \State $a_i \gets a_i + 1$
            \State $B \gets B - 1$
            \If{$B = 0$} \textbf{break}
            \EndIf
        \EndFor
    \EndWhile

    \State Restore original order for $\mathbf{A}$ using $\mathbf{I}$
    \State \Return $\mathbf{A}$
\EndFunction

\end{algorithmic}
\end{algorithm}

\begin{algorithm}[t!]
\footnotesize
\caption{In-Cluster Selection} \label{alg:InClusterSelectionModified}
\begin{algorithmic}[1]
\State \textbf{Input:} Cluster $c_i = \{I_1, I_2, \dots, I_{n_i}\}$, centroid $\mathbf{\mu}_i$, selection size $a_i$, weights $w_p$ and $w_d$. 
\State \textbf{Output:} Sub-Behavior Sequence $S_i^*$.

\Function{DynamicSelect}{$c_i, \mathbf{\mu}_i, a_i, w_p, w_d$}
    \State Initialize $c_i^* \gets \emptyset$  
    \State Compute item embeddings $\mathbf{E}(c_i) = \{\mathbf{e}_1, \mathbf{e}_2, \dots, \mathbf{e}_{n_i}\}$, where $\mathbf{e}_j$ is the embedding of item $I_j \in c_i$.  
    \State Select the initial item:
    \[
    \mathbf{e}_\text{init} = \arg \min_{\mathbf{e}_j \in \mathbf{E}(c_i)} d(\mathbf{e}_j, \mathbf{\mu}_i)
    \]
    \State Update $c_i^* \gets c_i^* \cup \{\mathbf{e}_\text{init}\}$ and $\mathbf{E}(c_i) \gets \mathbf{E}(c_i) \setminus \{\mathbf{e}_\text{init}\}$.  

    \While{$|c_i^*| < a_i$}
        \State \textbf{Compute Marginal Gains:}
        \ForAll{$\mathbf{e}_j \in \mathbf{E}(c_i)$}
            \State Compute prototypicality gain:
            \[
            g_p(\mathbf{e}_j) = \frac{w_p}{1 + d(\mathbf{e}_j, \mathbf{\mu}_i)}
            \]
            \State Compute diversity gain:
            \[
            g_d(\mathbf{e}_j) = \frac{2w_d}{c_i} \sum_{\mathbf{e}_b \in c_i^*} d(\mathbf{e}_j, \mathbf{e}_b)
            \]
        \EndFor
        \State \textbf{Evaluate Selection Priority:}
        \State Identify the item maximizing the combined gain:
        \[
        \mathbf{e}_j^* = \arg \max_{\mathbf{e}_j \in \mathbf{E}(c_i)} \left( g_p(\mathbf{e}_j) + g_d(\mathbf{e}_j) \right)
        \]
        \State Update $c_i^* \gets c_i^* \cup \{\mathbf{e}_j^*\}$ and $\mathbf{E}(c_i) \gets \mathbf{E}(c_i) \setminus \{\mathbf{e}_j^*\}$.  
    \EndWhile
    
    Chronologically sort $c_i^*$ to get $S_i^*$.
    \State \Return $S_i^*$
\EndFunction
\end{algorithmic}
\end{algorithm}

\begin{table*}[t]
\footnotesize
\centering
\caption{Time complexity analysis. Cluster, A.1 and A.2 refers to clustering method used in Section~\ref{sec: clustering}, Algorithms~\ref{alg:SamplingBudgetAllocation} and \ref{alg:InClusterSelectionModified}, respectively.}
\vspace{-2mm}
\label{tab:time_complexity_ana}
\begin{tabular}{ccc} 
\toprule
Agent LLM-UM Strategy & Offline Phase Complexity & Online Phase Complexity \\ 
\hline
$\text{AgentCF}_{\text{Recent + Reflection}}$    & –                              & $O(2kT + N_I \cdot T)$ \\[0.5em]
\hline
$\text{AgentCF}_{\text{Relevance + Reflection}}$   & $O(nd)$                       & $N_I \cdot O(2kT + d + T)$ \\[0.5em]\hline
$\text{Agent4Rec}_{\text{Recent + Summarization}}$   & –                              & $O(T + N_I \cdot T)$ \\[0.5em]\hline
$\text{Agent4Rec}_{\text{Relevance + Summarization}}$& $O(nd)$                       & $N_I \cdot O(d + 2T)$ \\[0.5em]\hline
$\text{AgentCF}_{\text{PersonaX}}$          & $O(C \cdot 2kT + nd + \text{Cluster} + \text{A.1} + \text{A.2})$ & $N_I \cdot O(T + d)$ \\[0.5em]\hline

$\text{Agent4Rec}_{\text{PersonaX}}$       & $O(CT + nd + \text{Cluster} + \text{A.1} + \text{A.2})$ & $N_I \cdot O(T + d)$ \\
\bottomrule
\end{tabular}
\end{table*}

\subsection{In-Cluster Selection}
\label{sec: in_cluster}

After partitioning user behaviors into semantically coherent clusters and each cluster is allocated with a sampling quota, we are to select a representative subset from each cluster. Data selection methods that greedily choose items closest to the cluster centroid (e.g., \cite{herding, ICARL, beyond}) yield overly homogeneous user profiling, while boundary-focused strategies (e.g., \cite{diet, forgetting_score}) risk overemphasizing diversity at the expense of prototypical patterns. To address these issues, we introduce a sampling strategy that balances prototypicality and diversity within each cluster. For a cluster $c_i$, its centroid is computed as $\mathbf{\mu}_i = \frac{1}{|c_i|} \sum_{I_j \in c_i} \mathbf{E}(I_j)$. Let $c_i^*$ denote the selected subset from $c_i$. Our goal is to maximize both the similarity of selected items to the centroid and the diversity among them:
  
{\footnotesize
$$
\begin{aligned}
    \max_{c_i^*} \Biggl( & w_p \cdot \sum_{I_j \in c_i^*} \frac{1}{1+d(\mathbf{e}_j, \mathbf{\mu}_i)} 
    + w_d \cdot \frac{2}{a_i} \sum_{\substack{I_a, I_b \in c_i^* \\ a \neq b}} d(\mathbf{e}_a,\mathbf{e}_b) \Biggr)
\end{aligned}
$$
}

Here, $w_p = \alpha^{-10}$ and $w_d = 1 - w_p$, with the hyperparameter $\alpha$ tuning the trade-off: values near 1.001 approximate centroid selection, while values around 1.4 approach boundary selection. Empirically, $\alpha$ is set between 1.06 and 1.08 (see Section~\ref{sec: hyper_parameter}). We frame the selection as discrete optimization problem and using a Greedy Selection algorithm (Algorithm~\ref{alg:InClusterSelectionModified}) to solve it, which iteratively selects the element with the highest marginal gain. A visual explanation of the selection algorithm is provided in Appendix~\ref{sec:app_in_cluster}, and Appendix~\ref{sec:theo_analysis} provides a convergence analysis of the proposed objective function and demonstrates how the greedy algorithm can attain suboptimal performance.

The design of our objective function is inspired by prior studies in data selection and empirical findings \cite{beyond}, which demonstrate that for small datasets, prioritizing simple, prototypical examples yields the greatest benefit, whereas for sufficiently large datasets, selection methods that emphasize harder examples improve the generalization of deep learning models.

\begin{figure}[h]
  \centering
  \vspace{-3mm}
    \includegraphics[width=2.9 in]{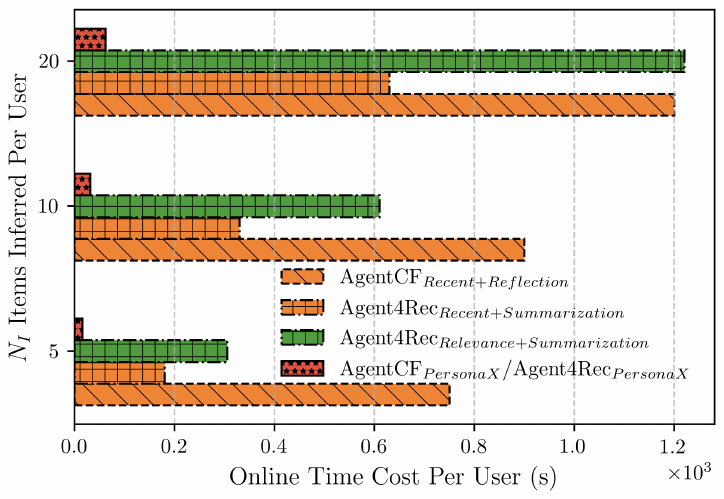}
    \vspace{-3mm}
   \caption{Online time cost analysis.}
    \vspace{-6mm}
    \label{fig: time_complexity}
\end{figure}

\subsection{Offline Profiling and Online Selection}
After selecting representative SBS, PersonaX continue to construct persona offline. Given a selected behavior subset $c_i^*$, we generate a corresponding persona $p_i = \mathcal{P}(c_i^*)$. To ensure contextually relevant recommendations, PersonaX retrieves the most pertinent persona snippet $P_{\text{selected}}$ online, which is integrated into prompt templates to instruct agent recommendation.

\section{Efficiency Analysis}
\label{sec: efficiency_analysis}
Recent sampling operates in constant time $O(1)$. Let $O(d)$ denotes the time required to encode an item into an embedding vector, and $O(n \log k)$ represents the complexity of selecting the $k$ most relevant/recent items from $n$ items. Thus Relevance sampling has a time complexity of $O(nd + n \log k)$. For SBS sampling applied in PersonaX, we use $O(\text{Cluster}+\text{Alg.1}+\text{Alg.2})$ represent the time cost for process we depict from Section~\ref{sec: clustering} to \ref{sec: in_cluster}. Let $C$ denote the number of clusters, $O(T)$ the time complexity of a single API request to the LLM, and $N_I$ the number of candidate items inferred per time. We perform an analysis of the time complexity during both offline and online stages associated with two recommendation agents AgentCF and Agent4Rec. Additionally, we evaluate their time cost with the use of PersonaX for comparison. The results are summarized in Table~\ref{tab:time_complexity_ana}. The detailed illustrations are provided in Appendix~\ref{app:time_complexity_analysis}.

The primary contributors to time consumption are $T$ and $d$, while $O(C)$, $O(n\log k)$, $O(\text{Cluster+A.1+A.2})$ in ranking, clustering, and sampling are negligible. Assuming $n = 500$, $C = 20$, $T = 3$, $d = 0.1$, and $k = 10$, and varying $N_I$ over 5, 10, and 20. Additionally, to model a realistic production setting in which recommendation agents server online continuously, we assume that persona in PersonaX is cached for $\mathcal{D}=10$ successive inference calls. In practice, this conservative threshold means the persona is refreshed whenever ten new user behaviors are observed, preventing profile staleness; the actual refresh interval can be tuned empirically. Because vanilla AgentCF and Agent4Rec must regenerate the user profile at every call, their online latency is multiplied by $\mathcal{D}$, whereas PersonaX-assisted variants avoid this overhead altogether. Figure \ref{fig: time_complexity} visualises the resulting online time consumption. The bar for AgentCF$_{\text{Relevance+Reflection}}$ is omitted because its latency is orders of magnitude higher than that of the other methods and would dominate the plot. We mainly make a comparison when backbone recommendation agent as Agent4Rec which is more realistic, we observe that PersonaX-assisted Agent4Rec reduces runtime by 95\% compared with the Agent4Rec variant that employs Relevance sampling, while recudes runtime by 91\% for variants that use Recent sampling.

\section{Experiments}
In this section, we are to address these research questions (\textbf{RQs}): 
$\bullet$ \textbf{RQ1}: How does PersonaX improve downstream agent recommendation, and how the performance compared with baseline approaches?
$\bullet$ \textbf{RQ2}: How does the sampling size of historical behaviors affect the efficacy of user modeling? 
$\bullet$ \textbf{RQ3:} How sensitive is our method to hyper-parameter settings, and how can optimal parameters be chosen?

\subsection{Experimental Setup}
\subsubsection{Datasets}
We evaluate on two widely used subsets of the Amazon review dataset \cite{amazon_review}: \textit{CDs and Vinyl} and \textit{Books}. For the CDs dataset, similar to the settings in \cite{AgentCF}, we consider two variants, $\texttt{CDs}_{\texttt{50}}$ and $\texttt{CDs}_{\texttt{200}}$, which have average user interaction sequence lengths of 50 and 200, respectively. For the Books dataset, rather than restricting each user’s interactions to 20 items as in \cite{Agent4Rec}, we adopt the approach outlined in \cite{MMM, search_based_UM} to construct longer sequences, resulting in $\texttt{Books}_{\texttt{480}}$. A more detailed description, statistical analysis, and reproducibility are provided in Appendix~\ref{sec:app_datasets}.

\subsubsection{Evaluation}
We utilize all the interaction data except the most recent one to construct the user's behavior history \cite{sasrec}. And the most recent interaction is reserved for testing. We randomly sample 9 negative items and combine them with the positive item, converting these 10 items into textual descriptions to form the candidate set. For evaluation metric, we adopt the typical top-$N$ metrics hit rate (HR@\{1, 5\}), normalized discounted cumulative gain (NDCG@\{5\})~\citep{NDCG} and Mean Reciprocal Rank (MRR@\{10\})~\cite{MRR}. For all evaluation metrics in our experiments, higher values indicate better performance. Also, an intuitive case study is provided in Appendix~\ref{app:app_case_study}.

\begin{table*}
\centering
\caption{Performance comparison study.}

\label{tab:comparison_results}
\resizebox{\textwidth}{!}{
\begin{tabular}{l|cccc|cccc|cccc} 
\hline
LLM-UM & \multicolumn{4}{c|}{Reflection}  & \multicolumn{8}{c}{Summarization} \\
\hline
Datasets  & \multicolumn{4}{c|}{$\texttt{CDs}_{\texttt{50}}$} & \multicolumn{4}{c|}{$\texttt{CDs}_{\texttt{200}}$} & \multicolumn{4}{c}{$\texttt{Books}_{\texttt{480}}$}  \\ 
\hline
Metrics   & Hit@1 & Hit@5 & NDCG@5 & MRR@10                & Hit@1 & Hit@5 & NDCG@5 & MRR@10                & Hit@1 & Hit@5 & NDCG@5 & MRR@10                  \\ 
\hline
Full     & 19.00 & 66.00 & 42.56  & 39.38                & 36.00 & 67.00 & 51.75  & 50.78                 & 19.00 & 50.00 & 34.59  & 35.76                   \\
Random    & 31.00 & 67.00 & 49.18  & 47.74                 & 36.00 & 68.00 & 51.26  & 50.24                 & 33.00 & 73.00 & 53.59  & 50.50                   \\
Recent    & 34.00 & 69.00 & 50.69  & 49.31                 & 39.00 & 68.00 & 53.89  & 53.34                 & 35.00 & 74.00 & 55.23  & 52.76                   \\
Relevance & 40.00 & 69.00 & 54.97  & 54.47                 & 51.00 & 73.00 & 61.73  & 61.98                 & 61.00 & 80.00 & 71.50  & 71.86                   \\
Centroid      & 43.00 & 66.00  & 55.21  & 55.91                 & 42.00  & 70.00  & 57.07   & 56.53                 & 60.00 & 81.00 & 71.61 & 70.67                    \\
Boundary   &  42.00  & 68.00 & 55.85  & 55.73                   & 48.00  & 66.00  & 57.13   & 58.71                & 58.00 & 80.00 & 70.38 & 69.55                    \\
Ours      & \color{purple}\textbf{45.00} & \color{purple}\textbf{72.00} & \color{purple}\textbf{57.34}  & \color{purple}\textbf{58.38}                 & \color{purple}\textbf{55.00} & \color{purple}\textbf{75.00} & \color{purple}\textbf{64.56}  & \color{purple}\textbf{65.06}                 & \color{purple}\textbf{65.00} & \color{purple}\textbf{83.00} & \color{purple}\textbf{74.26}  & \color{purple}\textbf{73.22}                   \\
\hline
\end{tabular}}
\vspace{-10pt}
\end{table*}

\subsubsection{Downstream Recommendation Agent}
We select two recommendation agents \textbf{AgentCF} \cite{AgentCF} which models user personas using a \underline{Reflection} mechanism, and \textbf{Agent4Rec} \cite{Agent4Rec} which captures users' unique preferences through a \underline{Summarization} method. Further details on the foundational methods can be found in Appendix~\ref{sec:app_backbone}. 
The original AgentCF offers two configurations—AgentCF (Recent+Reflection) and AgentCF (Relevance+Reflection)—while the standard Agent4Rec corresponds to Agent4Rec (Full+Summarization).

\subsubsection{Baseline Comparison}
To rigorously assess the benefit of integrating PersonaX, we enlarge the comparison scope beyond a mere juxtaposition of PersonaX-assisted AgentCF and Agent4Rec with their original implementations.  
Specifically, we pair two LLM-UM methods—Reflection and Summarization—with six representative behavior-sequence sampling strategies that serve as baselines:
(1) Full \cite{Agent4Rec}: Using complete user behavior sequence.
(2) Recent \cite{AgentCF}: Selecting the most recent behaviors to capture the user's short-term preferences.
(3) Relevance \cite{AgentCF,search_based_UM}: Retrieving the subset of behaviors most pertinent to the recommendation scenario from the user's long-term preferences.
(4) Random \cite{Deepcore,GDumb}: Randomly selecting a portion of behaviors, it is a robust and effective sampling method.
(5) Centroid Selection \cite{herding,ICARL,beyond}: As outlined in Section~\ref{sec: in_cluster}, we configure $\alpha=1.001$ in Algorithm~\ref{alg:InClusterSelectionModified}. This configuration prioritizes the selection of samples that are closest to the cluster centroid, effectively capturing the most prototypical data points within the cluster.  
(6) Boundary Selection \cite{diet,forgetting_score}: As detailed in Section~\ref{sec: in_cluster}, we set $\alpha=1.4$ in Algorithm~\ref{alg:InClusterSelectionModified}. Under this setting, the algorithm selects samples located at the cluster boundary and emphasizes the diversity coverage.

\subsubsection{Implementation Details}
We applied AgentCF to $\texttt{CDs}_{\texttt{50}}$, and Agent4Rec for $\texttt{CDs}_{\texttt{200}}$ and $\texttt{Books}_{\texttt{480}}$. For PersonaX, extensive experiments were conducted under diverse hyper-parameter configurations: the distance threshold $\tau \in \{0.5, 0.7\}$ and the trade-off parameter $\alpha \in \{1.01, 1.04, 1.08, 1.1\}$. Different selection ratios ($\frac{k}{n}$) were tested, including $\{10, 30, 50, 70, 90, 100\}$ for all three datasets. We also ensured that each cluster sampled at least one behavior by enforcing $k = \min(m, n \cdot \text{ratio})$. To evaluate the performance of the baseline methods, we varied the hyper-parameter selection ratio across different ranges for each dataset. Specifically, for $\texttt{CDs}_{\texttt{50}}$, the selection ratio was chosen from $\{0.02, 0.06, 0.08, 0.1, 0.16, 0.2, 0.3\}$. Similarly, for $\texttt{CDs}_{\texttt{200}}$, it ranged over $\{0.005, 0.01, 0.02, 0.03, 0.05, 0.08, 0.1\}$, and for $\texttt{Books}_{\texttt{480}}$, the selection ratio spanned $\{0.002, 0.005, 0.008, 0.011, 0.014\}$. These selection ratios settings were made to evaluate the baseline methods at equivalent levels of data resource utility, ensuring a fair and controlled comparison with PersonaX whose SBS sizes are listed in Table~\ref{tab:personaX_results_diff_ratio}. The prompt templates are provided in Appendix~\ref{sec:app_prompts}.

\begin{figure*}[t]
  \centering
    \includegraphics[width=6.2 in]{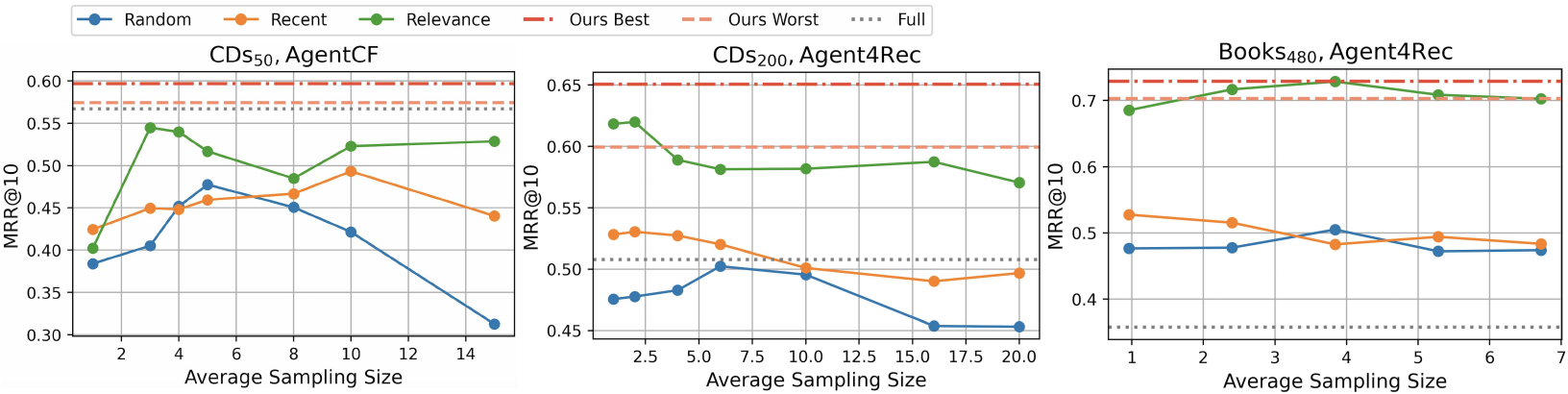}
    \vspace{-2mm}
   \caption{Analysis of the impact of sampling size on user modeling.}
   \vspace{-5mm}
    \label{fig: pre_study}
\end{figure*}

\subsection{Performance Evaluation (RQ 1)}
\vspace{-2mm}
Key observations and insights from Tables~\ref{tab:personaX_results_diff_ratio} highlight the robustness and effectiveness of our proposed method across various agent recommendation approaches, datasets, and evaluation metrics. PersonaX consistently outperforms the Full approach under any level of data resource utilization, even in scenarios where PersonaX achieves its least favorable results. Notably, on the $\texttt{Books}_{\texttt{480}}$ dataset, which features longer behavior sequences, our method achieves significant improvements over the Full methods. This phenomenon highlights the shortcomings of existing agent recommendation methods in handling long behavior sequences, but PersonaX fills this critical research gap.

Table~\ref{tab:comparison_results} reports the best MRR@10, highlighting PersonaX's performance advantages over baselines. Our approach demonstrates substantial improvements over the widely adopted and strong baseline method, Relevance. For example, on the $\texttt{CDs}_{\texttt{50}}$ dataset, our method achieves a Hit@1 score of 45.00, significantly exceeding the 40.00 obtained by Relevance. Similarly, we observe the suboptimal performance of the Centroid and Boundary methods, particularly on $\texttt{CDs}_{\texttt{200}}$. Upon analysis, we attribute the underperformance of the Centroid method to its tendency to sample overly homogeneous information, which results in overly simplistic and narrow user personas. While the Boundary method ensures sample diversity, an excessive focus on diversity can dilute the representation of typical user persona characteristics. In contrast, our method consistently delivers superior and stable performance, highlighting the effectiveness of balancing prototypicality and diversity. This equilibrium enables our approach to capture nuanced user personas with greater precision, establishing it as a robust and versatile solution for user modeling.


\begin{table}[t!]
\centering
\footnotesize
\caption{Performance of PersonaX at different selection ratios. We highlight \textcolor{purple}{\textbf{best performance}}, and the \textcolor{define_pink}{\textbf{worst performance}}.}
\label{tab:personaX_results_diff_ratio}
\vspace{-4mm}
\textbf{Reflection on $\texttt{CDs}_{\texttt{50}}$} \\[0.5em]
\begin{tabular}{l|l|llll}
\hline
Ratio  & \#SBS & HR@1                                 & HR@5                                 & NDCG@5                              & MRR                                \\ \hline
100    & 5.56  & 41.00                                & 67.00                                & \color{define_pink}\textbf{54.67}     & 54.67                             \\
90     & 4.69  & 42.00                                & 69.00                                & 55.66                               & 55.22                             \\
70     & 3.52  & \color{define_pink}\textbf{39.00}       & 70.00                                & 54.95                               & \color{define_pink}\textbf{53.50}    \\
50     & 2.88  & 41.00                                & 67.00                                & 54.69                               & 55.08                             \\
30     & 1.83  & \color{purple}\textbf{45.00}           & \color{purple}\textbf{72.00}           & \color{purple}\textbf{57.34}          & \color{purple}\textbf{58.38}         \\
10     & 1.0   & 42.00                                & \color{define_pink}\textbf{66.00}       & 56.07                               & 55.25                             \\ \hline
\end{tabular}

\vspace{0.5em}
\hrule\hrule
\vspace{0.5em}

\textbf{Summarization on $\texttt{CDs}_{\texttt{200}}$} \\[0.5em]
\begin{tabular}{l|l|llll}
\hline
Ratio  & \#SBS & HR@1                                 & HR@5                                 & NDCG@5                              & MRR                                \\ \hline
100    & 8.15  & \color{define_pink}\textbf{43.00}       & \color{define_pink}\textbf{68.00}       & \color{define_pink}\textbf{56.85}      & \color{define_pink}\textbf{57.07}     \\
90     & 7.19  & 49.00                                & 70.00                                & 59.66                               & 59.95                             \\
70     & 5.48  & 47.00                                & 71.00                                & 60.54                               & 60.54                             \\
50     & 3.59  & \color{purple}\textbf{55.00}           & \color{purple}\textbf{75.00}           & \color{purple}\textbf{64.56}          & \color{purple}\textbf{65.06}         \\
30     & 2.3   & 51.00                                & 73.00                                & 62.45                               & 62.42                             \\
10     & 1.0   & 47.00                                & 72.00                                & 61.91                               & 60.99                             \\ \hline
\end{tabular}

\vspace{0.5em}
\hrule\hrule
\vspace{0.5em}

\textbf{Summarization on $\texttt{Books}_{\texttt{480}}$} \\[0.5em]
\begin{tabular}{l|l|llll}
\hline
Ratio  & \#SBS & HR@1                                 & HR@5                                 & NDCG@5                              & MRR                                \\ \hline
100    & 15.35 & 61.00                                & 83.00                                & 73.56                               & 72.18                             \\
90     & 11.74 & \color{define_pink}\textbf{59.00}       & \color{define_pink}\textbf{80.00}       & \color{define_pink}\textbf{71.36}      & \color{define_pink}\textbf{71.70}     \\
70     & 8.41  & 64.00                                & 81.00                                & 72.55                               & 72.62                             \\
50     & 4.2   & \color{purple}\textbf{65.00}           & \color{purple}\textbf{83.00}           & \color{purple}\textbf{74.26}          & \color{purple}\textbf{73.22}         \\
30     & 1.82  & 64.00                                & 82.00                                & 73.68                               & 72.14                             \\
10     & 1.0   & 63.00                                & 83.00                                & 72.90                               & 71.75                             \\ \hline
\end{tabular}
\vspace{-7mm}
\end{table}

\subsection{Sampling Size Investigation (RQ 2)}
Understanding the influence of sequence length of SBS on the efficacy of user modeling is a pivotal research question. Traditional recommendation systems have largely relied on long-sequence modeling strategies, such as SIM \cite{search_based_UM}, which, when applied to datasets like Amazon Books, typically sample 10 interactions to approximate short-term behavioral patterns and 90 interactions for long-term modeling. However, in the context of LLM-UM, prior works such as AgentCF and Agent4Rec have yet to conduct a systematic investigation into the effect of sequence length on user modeling performance.

To address this gap, we first conduct analysis on PersonaX. As shown in Tables~\ref{tab:personaX_results_diff_ratio}, the results indicate that performance generally peaks at intermediate selection ratios or short SBS lengths. For instance, 30\% selection ratio for $\texttt{CDs}_{\texttt{50}}$ and 50\% for both $\texttt{CDs}_{\texttt{200}}$ and $\texttt{Books}_{\texttt{480}}$. We further examined the performance of three sampling strategies—Random, Recent, and Relevance—under varying sampling sizes, as illustrated in Figure~\ref{fig: pre_study}, finding that while initial increases in sampling size improve performance, oversampling eventually leads to performance deterioration. The optimal sampling size varies across datasets. Specifically, for the Relevance method, the ideal size is approximately 3, while the Recent method demonstrates heightened sensitivity to dataset characteristics, with the most recent single behavior often yielding strong results. For the Random method, a sampling size of around 5 is most effective.

\vspace{-2mm}
\subsection{Hyper-parameter Analysis (RQ3)}
\label{sec: hyper_parameter}
\begin{figure}[t]
  \centering
    \includegraphics[width=2.9 in]{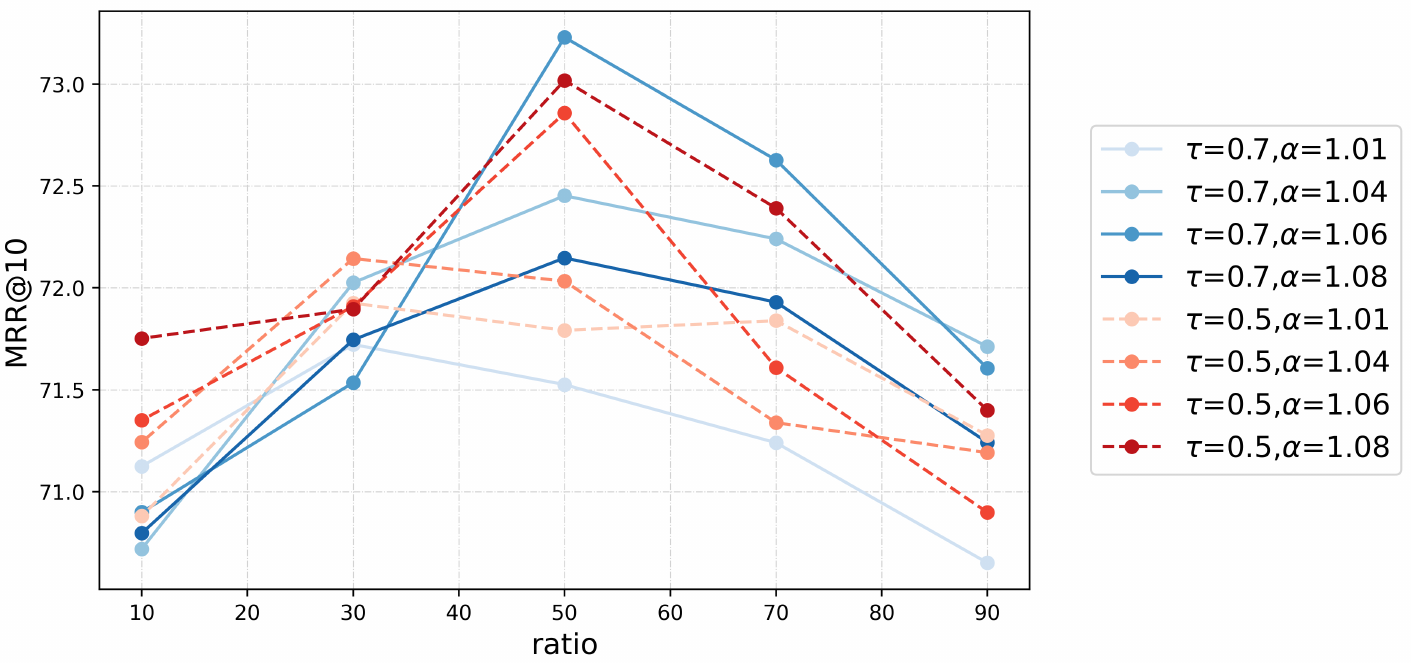}

   \caption{Impact of $\tau$ and $\alpha$ on PersonaX.}
    \vspace{-6mm}
    \label{fig: hyper_parameter}
\end{figure}

This section examines the impact of $\tau$ and $\alpha$ on PersonaX's performance. Our emperimental results, as illustrated in Figure~\ref{fig: hyper_parameter}, uncovers nuanced patterns in how these hyperparameters influence the model’s overall performance. In Appendix~\ref{sec:app_hyper}, we present more illustration alongside a visualization analysis of the sampling process.

The empirical results indicate that: (1) incorporating diverse samples is beneficial for enhancing performance. Specifically, higher values of $\alpha$ (e.g., 1.06–1.08) lead to significant performance improvements at large ratios (0.5–0.9); (2) PersonaX requires minimal fine-tuning effort within the range $\tau \in [0.5, 0.7], \alpha \in [1.04,1.08]$ and demonstrates robust performance, with a worst-case accuracy of 71.6, which closely approaches the best performance of the relevance baseline (71.86); (3) a higher $\tau$ expands the behavioral scope within clusters, making a lower $\alpha$ preferable to prevent excessive diversity. Conversely, a larger $\alpha$ prioritizes more diverse samples, necessitating a smaller $\tau$ to mitigate over-dispersion. For instance, when $\tau = 0.5$, a higher $\alpha$ (1.08) is appropriate, whereas $\tau = 0.7$ favors a slightly lower $\alpha$ (1.06) to avoid overemphasizing highly diverse samples.

\section{Related Works}
\subsection{Large Language Model for User Modeling}
User Modeling (UM) aims to extract valuable insights and patterns from users' long historical behavior sequence, and Large Language Models (LLMs) excel in characterizing user personalities and discerning preferences. Leveraging LLMs for UM has gained increasing attention, and the generated textual personas can be applied to downstream personalization tasks~\cite{slmrec,lightlm,nmcdr,amid}. For example, ONCE \cite{ONCE} utilizes ChatGPT to infer users' preferred topics and regions, enhancing click-through rate prediction with these generated profiles. Kang et al. \cite{kang2023llmsunderstanduserpreferences} enable LLMs to comprehend user preferences from behavior history to predict user ratings. LLMRec \cite{llmrec} identifies limitations in directly using raw item descriptions, which often fail to capture the subtle nuances of user preferences. To address this, it employs four distinct text enrichment strategies to enhance the input and improve recommendation performance. LLMRank \cite{llmrank} introduces specialized prompting and bootstrapping techniques that incorporate user interaction histories, effectively aligning with user intent. Moreover, two prominent strategies—Summarization and Reflection—have been widely adopted in leading agent recommendation frameworks, such as Agent4Rec \cite{Agent4Rec}, RecAgent \cite{RecAgent}, and AgentCF \cite{AgentCF}. Summarization focuses on distilling user behaviors, while reflection emphasizes iterative learning from interactions.

However, no research has focused on the performance of LLMs when handling extensive behaviors, nor has any LLM-UM method been proficient at efficiently and accurately modeling user personas from long behavior sequences. We are the first to address this gap and introduce PersonaX.

\subsection{Personalized Agents}  
LLM-driven agents have gained prominence for their autonomous decision-making, tool utilization \cite{Gpt4tools,Toolllm,a-mem,iagent}, and adaptive intelligence. Recent advances enable personalized agents through encoded personalities \cite{rao2023can}, backgrounds, and behavioral traits in prompts. Such persona-driven frameworks enhance user engagement through human-like interactions \cite{better_agent}, with applications like CharacterAgent \cite{character_agent} demonstrating consistent persona emulation of historical figures for immersive simulations. The personalization of agent also enable the simulations of social dynamics \cite{smallville}, competition \cite{CompeteAI}, and collaboration \cite{multiagent_colaboration}. 

However, recommendation agents \cite{rs_llm_agent_survey} face distinct challenges: Unlike predefined personas, user preferences in recommendation contexts are implicit and behaviorally embedded rather than verbally expressed. This creates alignment difficulties between agent decisions and users' latent preferences. The primary objective of PersonaX is to develop a highly accurate and realistic user modeling method, enabling instruction-based agents to consistently simulate and align with the decision-making behaviors of the users they surrogate.

\vspace{-2mm}
\section{Conclusion}
In this study, we present PersonaX, a LLM-UM framework oriented for agent recommendation specially designed for processing long behavior sequences. PersonaX utilizes only 30\%-50\% of the historical behavior data and strategically select high-quality sub-behavior sequences of short length (often $<5$) for generating broad spectrum of persona snippets offline. When PersonaX integrated into existing agent recommendation methods, such as AgentCF and Agent4Rec, PersonaX delivers substantial performance gains—ranging from 3\% to 11\% over AgentCF, and an impressive 10\% to 50\% improvement over Agent4Rec. Theoretical analysis indicates that integrating \textsc{PersonaX} into downstream recommendation agents markedly reduces online inference latency—a benefit that is especially pronounced in continuously servers. We believe that PersonaX significantly facilitate the agent recommendation in predictive accuracy and inference efficiency.

\section*{Limitations}
While PersonaX effectively tackles the challenge of modeling user behavior over extended sequences in LLM-based user modeling, its performance in real-world streaming data scenarios remains unexplored. This presents a promising opportunity for future enhancements.
A fundamental characteristic of PersonaX lies in its offline generation of multiple personas, capturing diverse aspects of user preferences. This design facilitates long-horizon modeling, where personas encapsulate user interests over extended periods and maintain their effectiveness for prolonged use, surpassing approaches (e.g., AgentCF) that depend on recent-sampling strategies and require frequent profile updates. However, an exciting direction for future work involves exploring the optimal duration for which these precomputed personas retain their efficacy in online deployment. Understanding the dynamics of performance degradation over time can inform strategies for adaptive persona updates. 

\section*{Ethics}
We use publicly available datasets collected under standard ethical protocols and strictly adhere to their intended research use. PersonaX is designed solely for academic purposes, and by following these safeguards, we uphold ethical standards in data usage, privacy protection, and transparency.

\section*{Acknowledgments}
This work was sponsored by the \texttt{Australian Research Council under the Linkage Projects Grant LP210100129}.



\bibliography{custom}

\begin{thebibliography}{56}
\providecommand{\natexlab}[1]{#1}

\bibitem[{Bai and Bilmes(2018)}]{greed_is_good}
Wenruo Bai and Jeff Bilmes. 2018.
\newblock \href {https://proceedings.mlr.press/v80/bai18a.html} {Greed is still
  good: Maximizing monotone submodular+supermodular (bp) functions}.
\newblock In \emph{Proceedings of the 35th International Conference on Machine
  Learning}, volume~80 of \emph{Proceedings of Machine Learning Research},
  pages 304--313. PMLR.

\bibitem[{Borgeaud et~al.(2022)Borgeaud, Mensch, Hoffmann, Cai, Rutherford,
  Millican, Van Den~Driessche, Lespiau, Damoc, Clark
  et~al.}]{improve_by_retrieve}
Sebastian Borgeaud, Arthur Mensch, Jordan Hoffmann, Trevor Cai, Eliza
  Rutherford, Katie Millican, George~Bm Van Den~Driessche, Jean-Baptiste
  Lespiau, Bogdan Damoc, Aidan Clark, et~al. 2022.
\newblock Improving language models by retrieving from trillions of tokens.
\newblock In \emph{International conference on machine learning}, pages
  2206--2240. PMLR.

\bibitem[{Chen et~al.(2024)Chen, Xiao, Zhang, Luo, Lian, and
  Liu}]{BGE_embedding}
Jianlyu Chen, Shitao Xiao, Peitian Zhang, Kun Luo, Defu Lian, and Zheng Liu.
  2024.
\newblock \href {https://aclanthology.org/2024.findings-acl.137}
  {{M}3-embedding: Multi-linguality, multi-functionality, multi-granularity
  text embeddings through self-knowledge distillation}.
\newblock In \emph{Findings of the Association for Computational Linguistics
  ACL 2024}, pages 2318--2335, Bangkok, Thailand and virtual meeting.
  Association for Computational Linguistics.

\bibitem[{Cheng et~al.(2023)Cheng, Luo, Chen, Liu, Zhao, and Yan}]{lift_up}
Xin Cheng, Di~Luo, Xiuying Chen, Lemao Liu, Dongyan Zhao, and Rui Yan. 2023.
\newblock \href {https://arxiv.org/abs/2305.02437} {Lift yourself up:
  Retrieval-augmented text generation with self memory}.
\newblock \emph{Preprint}, arXiv:2305.02437.

\bibitem[{Dai et~al.(2023)Dai, Shao, Zhao, Yu, Si, Xu, Sun, Zhang, and
  Xu}]{uncovering}
Sunhao Dai, Ninglu Shao, Haiyuan Zhao, Weijie Yu, Zihua Si, Chen Xu, Zhongxiang
  Sun, Xiao Zhang, and Jun Xu. 2023.
\newblock \href {https://doi.org/10.1145/3604915.3610646} {Uncovering
  chatgpt’s capabilities in recommender systems}.
\newblock In \emph{Proceedings of the 17th ACM Conference on Recommender
  Systems}, RecSys '23, page 1126–1132, New York, NY, USA. Association for
  Computing Machinery.

\bibitem[{Guo et~al.(2022)Guo, Zhao, and Bai}]{Deepcore}
Chengcheng Guo, Bo~Zhao, and Yanbing Bai. 2022.
\newblock Deepcore: A comprehensive library for coreset selection in deep
  learning.
\newblock In \emph{International Conference on Database and Expert Systems
  Applications}, pages 181--195. Springer.

\bibitem[{Hou et~al.(2024)Hou, Zhang, Lin, Lu, Xie, McAuley, and
  Zhao}]{llmrank}
Yupeng Hou, Junjie Zhang, Zihan Lin, Hongyu Lu, Ruobing Xie, Julian McAuley,
  and Wayne~Xin Zhao. 2024.
\newblock Large language models are zero-shot rankers for recommender systems.
\newblock In \emph{European Conference on Information Retrieval}, pages
  364--381. Springer.

\bibitem[{J{\"a}rvelin and Kek{\"a}l{\"a}inen(2002)}]{NDCG}
Kalervo J{\"a}rvelin and Jaana Kek{\"a}l{\"a}inen. 2002.
\newblock Cumulated gain-based evaluation of ir techniques.
\newblock \emph{ACM Transactions on Information Systems (TOIS)},
  20(4):422--446.

\bibitem[{Kang and McAuley(2018)}]{sasrec}
Wang-Cheng Kang and Julian McAuley. 2018.
\newblock Self-attentive sequential recommendation.
\newblock In \emph{2018 IEEE international conference on data mining (ICDM)},
  pages 197--206. IEEE.

\bibitem[{Kang et~al.(2023)Kang, Ni, Mehta, Sathiamoorthy, Hong, Chi, and
  Cheng}]{kang2023llmsunderstanduserpreferences}
Wang-Cheng Kang, Jianmo Ni, Nikhil Mehta, Maheswaran Sathiamoorthy, Lichan
  Hong, Ed~Chi, and Derek~Zhiyuan Cheng. 2023.
\newblock \href {https://arxiv.org/abs/2305.06474} {Do llms understand user
  preferences? evaluating llms on user rating prediction}.
\newblock \emph{Preprint}, arXiv:2305.06474.

\bibitem[{Lewis et~al.(2020)Lewis, Perez, Piktus, Petroni, Karpukhin, Goyal,
  K{\"u}ttler, Lewis, Yih, Rockt{\"a}schel et~al.}]{RAG}
Patrick Lewis, Ethan Perez, Aleksandra Piktus, Fabio Petroni, Vladimir
  Karpukhin, Naman Goyal, Heinrich K{\"u}ttler, Mike Lewis, Wen-tau Yih, Tim
  Rockt{\"a}schel, et~al. 2020.
\newblock Retrieval-augmented generation for knowledge-intensive nlp tasks.
\newblock \emph{Advances in Neural Information Processing Systems},
  33:9459--9474.

\bibitem[{Li et~al.(2023)Li, Al~Kader~Hammoud, Itani, Khizbullin, and
  Ghanem}]{CAMEL}
Guohao Li, Hasan~Abed Al~Kader~Hammoud, Hani Itani, Dmitrii Khizbullin, and
  Bernard Ghanem. 2023.
\newblock Camel: communicative agents for "mind" exploration of large language
  model society.
\newblock In \emph{Proceedings of the 37th International Conference on Neural
  Information Processing Systems}, NIPS '23, Red Hook, NY, USA. Curran
  Associates Inc.

\bibitem[{Liu et~al.(2023)Liu, Liu, Zhou, Lv, Zhou, and
  Zhang}]{chatgptgoodrecommender}
Junling Liu, Chao Liu, Peilin Zhou, Renjie Lv, Kang Zhou, and Yan Zhang. 2023.
\newblock \href {https://arxiv.org/abs/2304.10149} {Is chatgpt a good
  recommender? a preliminary study}.
\newblock \emph{Preprint}, arXiv:2304.10149.

\bibitem[{Liu et~al.(2024)Liu, Chen, Sakai, and Wu}]{ONCE}
Qijiong Liu, Nuo Chen, Tetsuya Sakai, and Xiao-Ming Wu. 2024.
\newblock \href {https://doi.org/10.1145/3616855.3635845} {Once: Boosting
  content-based recommendation with both open- and closed-source large language
  models}.
\newblock In \emph{Proceedings of the 17th ACM International Conference on Web
  Search and Data Mining}, WSDM '24, page 452–461, New York, NY, USA.
  Association for Computing Machinery.

\bibitem[{Luo et~al.(2023)Luo, He, Zhao, Huang, Zhou, Li, Xiao, Zhan, and
  Song}]{recranker}
Sichun Luo, Bowei He, Haohan Zhao, Yinya Huang, Aojun Zhou, Zongpeng Li,
  Yuanzhang Xiao, Mingjie Zhan, and Linqi Song. 2023.
\newblock Recranker: Instruction tuning large language model as ranker for
  top-k recommendation.
\newblock \emph{arXiv preprint arXiv:2312.16018}.

\bibitem[{Lyu et~al.(2024)Lyu, Jiang, Zeng, Xia, Wang, Zhang, Chen, Leung,
  Tang, and Luo}]{llmrec}
Hanjia Lyu, Song Jiang, Hanqing Zeng, Yinglong Xia, Qifan Wang, Si~Zhang, Ren
  Chen, Chris Leung, Jiajie Tang, and Jiebo Luo. 2024.
\newblock \href {https://doi.org/10.18653/v1/2024.findings-naacl.39}
  {{LLM}-rec: Personalized recommendation via prompting large language models}.
\newblock In \emph{Findings of the Association for Computational Linguistics:
  NAACL 2024}, pages 583--612, Mexico City, Mexico. Association for
  Computational Linguistics.

\bibitem[{Mei and Zhang(2023)}]{lightlm}
Kai Mei and Yongfeng Zhang. 2023.
\newblock Lightlm: a lightweight deep and narrow language model for generative
  recommendation.
\newblock \emph{arXiv preprint arXiv:2310.17488}.

\bibitem[{Ni et~al.(2019)Ni, Li, and McAuley}]{amazon_review}
Jianmo Ni, Jiacheng Li, and Julian McAuley. 2019.
\newblock Justifying recommendations using distantly-labeled reviews and
  fine-grained aspects.
\newblock In \emph{Proceedings of the 2019 conference on empirical methods in
  natural language processing and the 9th international joint conference on
  natural language processing (EMNLP-IJCNLP)}, pages 188--197.

\bibitem[{Park et~al.(2023)Park, O'Brien, Cai, Morris, Liang, and
  Bernstein}]{smallville}
Joon~Sung Park, Joseph O'Brien, Carrie~Jun Cai, Meredith~Ringel Morris, Percy
  Liang, and Michael~S Bernstein. 2023.
\newblock Generative agents: Interactive simulacra of human behavior.
\newblock In \emph{Proceedings of the 36th annual acm symposium on user
  interface software and technology}, pages 1--22.

\bibitem[{Paul et~al.(2021)Paul, Ganguli, and Dziugaite}]{diet}
Mansheej Paul, Surya Ganguli, and Gintare~Karolina Dziugaite. 2021.
\newblock Deep learning on a data diet: Finding important examples early in
  training.
\newblock \emph{Advances in Neural Information Processing Systems},
  34:20596--20607.

\bibitem[{Pi et~al.(2019)Pi, Bian, Zhou, Zhu, and Gai}]{MMM}
Qi~Pi, Weijie Bian, Guorui Zhou, Xiaoqiang Zhu, and Kun Gai. 2019.
\newblock \href {https://doi.org/10.1145/3292500.3330666} {Practice on long
  sequential user behavior modeling for click-through rate prediction}.
\newblock In \emph{Proceedings of the 25th ACM SIGKDD International Conference
  on Knowledge Discovery \& Data Mining}, KDD '19, page 2671–2679, New York,
  NY, USA. Association for Computing Machinery.

\bibitem[{Pi et~al.(2020)Pi, Zhou, Zhang, Wang, Ren, Fan, Zhu, and
  Gai}]{search_based_UM}
Qi~Pi, Guorui Zhou, Yujing Zhang, Zhe Wang, Lejian Ren, Ying Fan, Xiaoqiang
  Zhu, and Kun Gai. 2020.
\newblock \href {https://doi.org/10.1145/3340531.3412744} {Search-based user
  interest modeling with lifelong sequential behavior data for click-through
  rate prediction}.
\newblock In \emph{Proceedings of the 29th ACM International Conference on
  Information \& Knowledge Management}, CIKM '20, page 2685–2692, New York,
  NY, USA. Association for Computing Machinery.

\bibitem[{Prabhu et~al.(2020)Prabhu, Torr, and Dokania}]{GDumb}
Ameya Prabhu, Philip H.~S. Torr, and Puneet~K. Dokania. 2020.
\newblock \href {https://doi.org/10.1007/978-3-030-58536-5_31} {Gdumb: A simple
  approach that questions our progress in continual learning}.
\newblock In \emph{Computer Vision – ECCV 2020: 16th European Conference,
  Glasgow, UK, August 23–28, 2020, Proceedings, Part II}, page 524–540,
  Berlin, Heidelberg. Springer-Verlag.

\bibitem[{Qin et~al.(2023)Qin, Liang, Ye, Zhu, Yan, Lu, Lin, Cong, Tang, Qian,
  Zhao, Hong, Tian, Xie, Zhou, Gerstein, Li, Liu, and Sun}]{Toolllm}
Yujia Qin, Shihao Liang, Yining Ye, Kunlun Zhu, Lan Yan, Yaxi Lu, Yankai Lin,
  Xin Cong, Xiangru Tang, Bill Qian, Sihan Zhao, Lauren Hong, Runchu Tian,
  Ruobing Xie, Jie Zhou, Mark Gerstein, Dahai Li, Zhiyuan Liu, and Maosong Sun.
  2023.
\newblock \href {https://arxiv.org/abs/2307.16789} {Toolllm: Facilitating large
  language models to master 16000+ real-world apis}.
\newblock \emph{Preprint}, arXiv:2307.16789.

\bibitem[{Rao et~al.(2023)Rao, Leung, and Miao}]{rao2023can}
Haocong Rao, Cyril Leung, and Chunyan Miao. 2023.
\newblock Can {ChatGPT} assess human personalities? a general evaluation
  framework.
\newblock \emph{arXiv preprint arXiv:2303.01248}.

\bibitem[{Rebuffi et~al.(2017)Rebuffi, Kolesnikov, Sperl, and Lampert}]{ICARL}
Sylvestre-Alvise Rebuffi, Alexander Kolesnikov, Georg Sperl, and Christoph~H
  Lampert. 2017.
\newblock icarl: Incremental classifier and representation learning.
\newblock In \emph{Proceedings of the IEEE conference on Computer Vision and
  Pattern Recognition}, pages 2001--2010.

\bibitem[{Ren and Huang(2024)}]{EasyRec}
Xubin Ren and Chao Huang. 2024.
\newblock \href {https://arxiv.org/abs/2408.08821} {Easyrec: Simple yet
  effective language models for recommendation}.
\newblock \emph{Preprint}, arXiv:2408.08821.

\bibitem[{Richardson et~al.(2023)Richardson, Zhang, Gillespie, Kar, Singh,
  Raeesy, Khan, and Sethy}]{Summarization}
Chris Richardson, Yao Zhang, Kellen Gillespie, Sudipta Kar, Arshdeep Singh,
  Zeynab Raeesy, Omar~Zia Khan, and Abhinav Sethy. 2023.
\newblock \href
  {https://www.amazon.science/publications/integrating-summarization-and-retrieval-for-enhanced-personalization-via-large-language-models}
  {Integrating summarization and retrieval for enhanced personalization via
  large language models}.
\newblock In \emph{Proceedings of the First Workshop on Personalized Generative
  AI (PGAI '23), co-located with the 32nd ACM International Conference on
  Information and Knowledge Management (CIKM 2023)}, Birmingham, United
  Kingdom. Association for Computing Machinery.
\newblock Workshop Paper.

\bibitem[{Salemi et~al.(2024)Salemi, Mysore, Bendersky, and Zamani}]{lamp}
Alireza Salemi, Sheshera Mysore, Michael Bendersky, and Hamed Zamani. 2024.
\newblock \href {https://doi.org/10.18653/v1/2024.acl-long.399} {{L}a{MP}: When
  large language models meet personalization}.
\newblock In \emph{Proceedings of the 62nd Annual Meeting of the Association
  for Computational Linguistics (Volume 1: Long Papers)}, pages 7370--7392,
  Bangkok, Thailand. Association for Computational Linguistics.

\bibitem[{Sarwar et~al.(2001)Sarwar, Karypis, Konstan, and Riedl}]{MRR}
Badrul Sarwar, George Karypis, Joseph Konstan, and John Riedl. 2001.
\newblock Item-based collaborative filtering recommendation algorithms.
\newblock In \emph{Proceedings of the 10th international conference on World
  Wide Web}, pages 285--295.

\bibitem[{Shao et~al.(2023)Shao, Li, Dai, and Qiu}]{character_agent}
Yunfan Shao, Linyang Li, Junqi Dai, and Xipeng Qiu. 2023.
\newblock \href {https://aclanthology.org/2023.emnlp-main.814}
  {Character-{LLM}: A trainable agent for role-playing}.
\newblock In \emph{Proceedings of the 2023 Conference on Empirical Methods in
  Natural Language Processing}, pages 13153--13187, Singapore. Association for
  Computational Linguistics.

\bibitem[{Shi et~al.(2024{\natexlab{a}})Shi, Xu, Zhang, Zi, and Wu}]{ACN}
Yunxiao Shi, Min Xu, Haimin Zhang, Xing Zi, and Qiang Wu. 2024{\natexlab{a}}.
\newblock \href {https://arxiv.org/abs/2409.00636} {A learnable agent
  collaboration network framework for personalized multimodal ai search
  engine}.
\newblock \emph{Preprint}, arXiv:2409.00636.

\bibitem[{Shi et~al.(2024{\natexlab{b}})Shi, Zi, Shi, Zhang, Wu, and
  Xu}]{Shi2024}
Yunxiao Shi, Xing Zi, Zijing Shi, Haimin Zhang, Qiang Wu, and Min Xu.
  2024{\natexlab{b}}.
\newblock \href {https://doi.org/10.3233/FAIA240748} {Enhancing retrieval and
  managing retrieval: A four-module synergy for improved quality and efficiency
  in rag systems}.
\newblock In \emph{ECAI 2024}, pages 2258--2265. IOS Press.

\bibitem[{Shi et~al.(2024{\natexlab{c}})Shi, Zi, Shi, Zhang, Wu, and
  Xu}]{ERAGent}
Yunxiao Shi, Xing Zi, Zijing Shi, Haimin Zhang, Qiang Wu, and Min Xu.
  2024{\natexlab{c}}.
\newblock \href {https://arxiv.org/abs/2405.06683} {Eragent: Enhancing
  retrieval-augmented language models with improved accuracy, efficiency, and
  personalization}.
\newblock \emph{Preprint}, arXiv:2405.06683.

\bibitem[{Sorscher et~al.(2022)Sorscher, Geirhos, Shekhar, Ganguli, and
  Morcos}]{beyond}
Ben Sorscher, Robert Geirhos, Shashank Shekhar, Surya Ganguli, and Ari~S
  Morcos. 2022.
\newblock Beyond neural scaling laws: beating power law scaling via data
  pruning.
\newblock \emph{arXiv preprint arXiv:2206.14486}.

\bibitem[{Sun et~al.(2024)Sun, Zhan, and Such}]{better_agent}
Guangzhi Sun, Xiao Zhan, and Jose Such. 2024.
\newblock \href {https://doi.org/10.1145/3640794.3665887} {Building better ai
  agents: A provocation on the utilisation of persona in llm-based
  conversational agents}.
\newblock In \emph{Proceedings of the 6th ACM Conference on Conversational User
  Interfaces}, CUI '24, New York, NY, USA. Association for Computing Machinery.

\bibitem[{Toneva et~al.(2019)Toneva, Sordoni, des Combes, Trischler, Bengio,
  and Gordon}]{forgetting_score}
Mariya Toneva, Alessandro Sordoni, Remi~Tachet des Combes, Adam Trischler,
  Yoshua Bengio, and Geoffrey~J. Gordon. 2019.
\newblock \href {https://arxiv.org/abs/1812.05159} {An empirical study of
  example forgetting during deep neural network learning}.
\newblock \emph{Preprint}, arXiv:1812.05159.

\bibitem[{Tran et~al.(2025)Tran, Dao, Nguyen, Pham, O'Sullivan, and
  Nguyen}]{multiagent_colaboration}
Khanh-Tung Tran, Dung Dao, Minh-Duong Nguyen, Quoc-Viet Pham, Barry O'Sullivan,
  and Hoang~D. Nguyen. 2025.
\newblock \href {https://arxiv.org/abs/2501.06322} {Multi-agent collaboration
  mechanisms: A survey of llms}.
\newblock \emph{Preprint}, arXiv:2501.06322.

\bibitem[{Wang et~al.(2024{\natexlab{a}})Wang, Ma, Feng, Zhang, Yang, Zhang,
  Chen, Tang, Chen, Lin et~al.}]{wang2024survey}
Lei Wang, Chen Ma, Xueyang Feng, Zeyu Zhang, Hao Yang, Jingsen Zhang, Zhiyuan
  Chen, Jiakai Tang, Xu~Chen, Yankai Lin, et~al. 2024{\natexlab{a}}.
\newblock A survey on large language model based autonomous agents.
\newblock \emph{Frontiers of Computer Science}, 18(6):186345.

\bibitem[{Wang et~al.(2024{\natexlab{b}})Wang, Zhang, Yang, Chen, Tang, Zhang,
  Chen, Lin, Song, Zhao, Xu, Dou, Wang, and Wen}]{RecAgent}
Lei Wang, Jingsen Zhang, Hao Yang, Zhiyuan Chen, Jiakai Tang, Zeyu Zhang,
  Xu~Chen, Yankai Lin, Ruihua Song, Wayne~Xin Zhao, Jun Xu, Zhicheng Dou, Jun
  Wang, and Ji-Rong Wen. 2024{\natexlab{b}}.
\newblock \href {https://arxiv.org/abs/2306.02552} {User behavior simulation
  with large language model based agents}.
\newblock \emph{Preprint}, arXiv:2306.02552.

\bibitem[{Wang et~al.(2024{\natexlab{c}})Wang, Liu, Zhang, Zhong, Wang, Yin,
  Fan, Wu, and Wen}]{RCAgent}
Zefan Wang, Zichuan Liu, Yingying Zhang, Aoxiao Zhong, Jihong Wang, Fengbin
  Yin, Lunting Fan, Lingfei Wu, and Qingsong Wen. 2024{\natexlab{c}}.
\newblock \href {https://doi.org/10.1145/3627673.3680016} {Rcagent: Cloud root
  cause analysis by autonomous agents with tool-augmented large language
  models}.
\newblock In \emph{Proceedings of the 33rd ACM International Conference on
  Information and Knowledge Management}, CIKM '24, page 4966–4974, New York,
  NY, USA. Association for Computing Machinery.

\bibitem[{Welling(2009)}]{herding}
Max Welling. 2009.
\newblock Herding dynamical weights to learn.
\newblock In \emph{Proceedings of the 26th Annual International Conference on
  Machine Learning}, pages 1121--1128.

\bibitem[{Xu et~al.(2023)Xu, Li, Ha, Guo, Ma, Liu, Chen, and Zhu}]{nmcdr}
Wujiang Xu, Shaoshuai Li, Mingming Ha, Xiaobo Guo, Qiongxu Ma, Xiaolei Liu,
  Linxun Chen, and Zhenfeng Zhu. 2023.
\newblock Neural node matching for multi-target cross domain recommendation.
\newblock In \emph{2023 IEEE 39th International Conference on Data Engineering
  (ICDE)}, pages 2154--2166. IEEE.

\bibitem[{Xu et~al.(2025{\natexlab{a}})Xu, Liang, Mei, Gao, Tan, and
  Zhang}]{a-mem}
Wujiang Xu, Zujie Liang, Kai Mei, Hang Gao, Juntao Tan, and Yongfeng Zhang.
  2025{\natexlab{a}}.
\newblock A-mem: Agentic memory for llm agents.
\newblock \emph{arXiv preprint arXiv:2502.12110}.

\bibitem[{Xu et~al.(2025{\natexlab{b}})Xu, Shi, Liang, Ning, Mei, Wang, Zhu,
  Xu, and Zhang}]{iagent}
Wujiang Xu, Yunxiao Shi, Zujie Liang, Xuying Ning, Kai Mei, Kun Wang, Xi~Zhu,
  Min Xu, and Yongfeng Zhang. 2025{\natexlab{b}}.
\newblock Instructagent: Building user controllable recommender via llm agent.
\newblock \emph{arXiv preprint arXiv:2502.14662}.

\bibitem[{Xu et~al.(2024{\natexlab{a}})Xu, Wu, Liang, Han, Ning, Shi, Lin, and
  Zhang}]{slmrec}
Wujiang Xu, Qitian Wu, Zujie Liang, Jiaojiao Han, Xuying Ning, Yunxiao Shi,
  Wenfang Lin, and Yongfeng Zhang. 2024{\natexlab{a}}.
\newblock Slmrec: empowering small language models for sequential
  recommendation.
\newblock In \emph{The Thirteenth International Conference on Learning
  Representations}.

\bibitem[{Xu et~al.(2024{\natexlab{b}})Xu, Wu, Wang, Ha, Ma, Chen, Han, and
  Yan}]{amid}
Wujiang Xu, Qitian Wu, Runzhong Wang, Mingming Ha, Qiongxu Ma, Linxun Chen,
  Bing Han, and Junchi Yan. 2024{\natexlab{b}}.
\newblock Rethinking cross-domain sequential recommendation under open-world
  assumptions.
\newblock In \emph{Proceedings of the ACM Web Conference 2024}, pages
  3173--3184.

\bibitem[{Yang et~al.(2023)Yang, Song, Li, Zhao, Ge, Li, and Shan}]{Gpt4tools}
Rui Yang, Lin Song, Yanwei Li, Sijie Zhao, Yixiao Ge, Xiu Li, and Ying Shan.
  2023.
\newblock \href {https://arxiv.org/abs/2305.18752} {Gpt4tools: Teaching large
  language model to use tools via self-instruction}.
\newblock \emph{Preprint}, arXiv:2305.18752.

\bibitem[{Zhang et~al.(2024{\natexlab{a}})Zhang, Chen, Sheng, Wang, and
  Chua}]{Agent4Rec}
An~Zhang, Yuxin Chen, Leheng Sheng, Xiang Wang, and Tat-Seng Chua.
  2024{\natexlab{a}}.
\newblock \href {https://doi.org/10.1145/3626772.3657844} {On generative agents
  in recommendation}.
\newblock In \emph{Proceedings of the 47th International ACM SIGIR Conference
  on Research and Development in Information Retrieval}, SIGIR '24, page
  1807–1817, New York, NY, USA. Association for Computing Machinery.

\bibitem[{Zhang et~al.(2024{\natexlab{b}})Zhang, Hou, Xie, Sun, McAuley, Zhao,
  Lin, and Wen}]{AgentCF}
Junjie Zhang, Yupeng Hou, Ruobing Xie, Wenqi Sun, Julian McAuley, Wayne~Xin
  Zhao, Leyu Lin, and Ji-Rong Wen. 2024{\natexlab{b}}.
\newblock \href {https://doi.org/10.1145/3589334.3645537} {Agentcf:
  Collaborative learning with autonomous language agents for recommender
  systems}.
\newblock In \emph{Proceedings of the ACM Web Conference 2024}, WWW '24, page
  3679–3689, New York, NY, USA. Association for Computing Machinery.

\bibitem[{Zhao et~al.(2024{\natexlab{a}})Zhao, Huang, Xu, Lin, Liu, and
  Huang}]{ExpeL}
Andrew Zhao, Daniel Huang, Quentin Xu, Matthieu Lin, Yong-Jin Liu, and Gao
  Huang. 2024{\natexlab{a}}.
\newblock \href {https://doi.org/10.1609/aaai.v38i17.29936} {Expel: Llm agents
  are experiential learners}.
\newblock In \emph{Thirty-Eighth {AAAI} Conference on Artificial Intelligence,
  {AAAI} 2024, Thirty-Sixth Conference on Innovative Applications of Artificial
  Intelligence, {IAAI} 2024, Fourteenth Symposium on Educational Advances in
  Artificial Intelligence, {EAAI} 2024, February 20-27, 2024, Vancouver,
  Canada}, pages 19632--19642. {AAAI} Press.

\bibitem[{Zhao et~al.(2024{\natexlab{b}})Zhao, Wang, Cen, Zha, Tan, Dong, and
  Tang}]{LongRAG}
Qingfei Zhao, Ruobing Wang, Yukuo Cen, Daren Zha, Shicheng Tan, Yuxiao Dong,
  and Jie Tang. 2024{\natexlab{b}}.
\newblock \href {https://arxiv.org/abs/2410.18050} {Longrag: A dual-perspective
  retrieval-augmented generation paradigm for long-context question answering}.
\newblock \emph{Preprint}, arXiv:2410.18050.

\bibitem[{Zhao et~al.()Zhao, Wang, Zhang, Jin, Zhu, Chen, and Xie}]{CompeteAI}
Qinlin Zhao, Jindong Wang, Yixuan Zhang, Yiqiao Jin, Kaijie Zhu, Hao Chen, and
  Xing Xie.
\newblock Competeai: Understanding the competition dynamics of large language
  model-based agents.
\newblock In \emph{Forty-first International Conference on Machine Learning}.

\bibitem[{Zheng et~al.(2023)Zheng, Liu, Lai, and Prakash}]{CCS}
Haizhong Zheng, Rui Liu, Fan Lai, and Atul Prakash. 2023.
\newblock \href {https://arxiv.org/abs/2210.15809} {Coverage-centric coreset
  selection for high pruning rates}.
\newblock \emph{Preprint}, arXiv:2210.15809.

\bibitem[{Zhou et~al.(2024)Zhou, Zhu, Jin, and
  Dou}]{personalized_search_memory}
Yujia Zhou, Qiannan Zhu, Jiajie Jin, and Zhicheng Dou. 2024.
\newblock \href {https://doi.org/10.1145/3589334.3645482} {Cognitive
  personalized search integrating large language models with an efficient
  memory mechanism}.
\newblock In \emph{Proceedings of the ACM Web Conference 2024}, WWW '24, page
  1464–1473, New York, NY, USA. Association for Computing Machinery.

\bibitem[{Zhu et~al.(2025)Zhu, Wang, Gao, Xu, Wang, Liu, Wang, Jin, Pang, Weng,
  Yu, and Zhang}]{rs_llm_agent_survey}
Xi~Zhu, Yu~Wang, Hang Gao, Wujiang Xu, Chen Wang, Zhiwei Liu, Kun Wang, Mingyu
  Jin, Linsey Pang, Qingsong Weng, Philip~S. Yu, and Yongfeng Zhang. 2025.
\newblock \href {https://doi.org/10.1561/3300000050} {Recommender systems meet
  large language model agents: A survey}.
\newblock \emph{Foundations and Trends® in Privacy and Security},
  7(4):247--396.

\end{thebibliography}


\clearpage
\onecolumn
\tableofcontents

\section*{APPENDIX}
\appendix

\section{Time Complexity Analysis}
\label{app:time_complexity_analysis}
In this section, we provide a rigorous analysis of the time complexity of user modeling in AgentCF and Agent4Rec, and examine how these complexities change when each method leverages PersonaX.

\subsection{Preliminary Analysis}
We begin by analyzing the time complexity of two sampling approaches. \textbf{Recent sampling}: This approach selects the most recent $k$ user behaviors, which requires $O(1)$ computational complexity. Let $\mathcal{F}$ be the hardware’s floating-point throughput, hence Recent sampling's time complexity in seconds is $O(1/\mathcal{F})$.
\textbf{Relevance Strategy}: This strategy identifies user behaviors most pertinent to the target item $I_{\text{target}}$. Encoding an item into a feature vector using a large language embedding model incurs a time complexity of $O(d)$, and thus encoding $n$ items results in a complexity of $O(nd)$. Selecting the top $k$ most relevant items requires $O(n \log k / \mathcal{F})$, leading to an overall complexity of $O(nd + n \log k/ \mathcal{F})$.

\subsection{Analysis of LLM-UM methods}

\textbf{Summarization}: In this method, the short behavioral sequence (SBS) $\mathcal{S}^*$ is distilled into a user representation $\mathcal{P}(\mathcal{S})$, and its complexity is independent of the sequence length $k$. Thus, the overall complexity remains $O(T)$.
\textbf{Reflection}: This method iteratively updates the user persona along with $\mathcal{S}^*$. In the best case, all inferences are correct on the first attempt, incurring a complexity of $O(kT)$. In the worst case, all initial inferences fail, requiring a single reflection to update the persona, which enables the second inference to be correct. This results in a complexity of $O(3kT)$. Taking an average across these cases yields an approximate complexity of $O(2kT)$. Given that commonly $k\leq 10$, this constant factor remains manageable.

\subsection{Analysis for AgentCF and Agent4Rec}
\textbf{AgentCF (Recent + Reflection)}. As this method updates the user profile dynamically with new behaviors, it is not well suited for offline profiling and be cached for long-term usage. The user profile is constructed once with a complexity of $O(1) + O(2kT)$. The profile is reused for inferring $N_I$ items, each requiring $O(T)$. Thus the overall online complexity is $O(1/ \mathcal{F} + 2kT + N_I T)$.

\noindent \textbf{AgentCF (Relevance + Reflection)}. Each item in the user's behavior sequence is embedded with a complexity of $O(nd)$. The user profile is constructed with a complexity of $O(n \log k) + O(2kT)$. For each inference, the complexity is $O(d + T)$. Thus the overall online complexity is $N_I \cdot O(n \log k/ \mathcal{F} + 2kT + d + T)$.

\noindent \textbf{Agent4Rec (Relevance + Summarization)}. Each item is embedded with a complexity of $O(nd)$. The user profile is constructed once with a complexity of $O(n \log k) + O(T)$. Each inference has a complexity of $O(d + T)$. Thus the overall online complexity is $N_I \cdot O(n \log k/ \mathcal{F} + d + 2T)$.

\noindent \textbf{Agent4Rec (Recent + Summarization)}. As this method updates the user profile dynamically, it is not suited for offline profiling. The user profile is constructed once with a complexity of $O(1) + O(T)$. Thus the overall online complexity is $O(1/ \mathcal{F} + T + N_I T)$.

\noindent \textbf{Agent4Rec+PersonaX}. Item embedding incurs $O(nd)$. The sampling process has a complexity of $O(\text{Cluster} + A.1 + A.2)$. Multiple persona generation requires $O(C T)$. The overall offline complexity is $O(C T + nd + \text{Cluster} + A.1 + A.2)$. For online phase, retrieving the user profile incurs $O(d)+O(1)$, and each inference requires $O(T)$. Thus the overall online complexity is $N_I \cdot O(T + 1/ \mathcal{F} + d)$.

\noindent \textbf{AgentCF+PersonaX}. Item embedding incurs $O(nd)$. The sampling process has a complexity of $O(\text{Cluster} + A.1 + A.2)$. Multiple persona generation requires $O(C \cdot 2kT)$. The overall offline complexity is $O(C \cdot 2kT + nd + \text{Cluster} + A.1 + A.2)$. For online phase, retrieving the user profile incurs $O(d) + O(1)$, and each inference requires $O(T)$. Thus the overall online complexity is $N_I \cdot O(T + 1/ \mathcal{F} + d)$.

 Since the terms $O(1/\mathcal{F})$ and $O(n\log k/\mathcal{F})$ are asymptotically negligible compared to $O(T)$, they can be safely omitted. Thus we can get the results presented in Table~\ref{tab:time_complexity_ana}.

\begin{table*}[ht]
\footnotesize
\centering
\caption{Summary of preprocessed subset statistics. "Avg.L" represents the average length of user behavior sequences.}

\label{tab:dataset}
\begin{tabular}{llllll} 
\toprule
\textbf{Subsets}       & \textbf{\textbf{\#Users}} & \textbf{\textbf{\#Items}} & \textbf{\textbf{\#Inters}} & \textbf{Sparsity} & \textbf{Avg.L}  \\ 
\hline
  $\texttt{CDs}_{\texttt{50}}$    &  100            & 4,899                       & 5,000                        & 98.97\%           & 50.00            \\ 
$\texttt{CDs}_{\texttt{200}}$  &  1000           & 101,902                       &    200,336                        &    99.80\%               &   200.34              \\
$\texttt{Books}_{\texttt{480}}$    &  1000              & 222,539                       &   481,455                         &     99.78\%              &  481.46               \\ 
\bottomrule
\end{tabular}
\end{table*}

\section{Datasets}
\label{sec:app_datasets}
In this appendix, we provide a detailed description of the dataset construction and statistics.

Building on prior studies such as AgentCF \cite{AgentCF}, Agent4Rec \cite{Agent4Rec}, and EasyRec \cite{EasyRec}, we evaluate our proposed method using two widely adopted subsets of the Amazon review dataset \cite{amazon_review}: \textit{CDs and Vinyl} and \textit{Books}. For the $\texttt{CDs}$ dataset, we construct $\texttt{CDs}_\texttt{50}$, and $\texttt{CDs}_\texttt{200}$, with average user interaction sequence lengths of 50 and 200, respectively. These settings are similar as those used in AgentCF \cite{AgentCF}. 

For the $\texttt{Books}$ dataset, departing from the approach of Agent4Rec which limits each user’s interactions to 20 items, we follow the guidelines of \cite{MMM, search_based_UM} to construct longer interaction sequences. Specifically, we create $\texttt{Books}_\texttt{480}$, with average sequence lengths of 480, respectively. Detailed statistics for these datasets are provided in Table \ref{tab:dataset}. 

Due to the high computational cost and expense associated with API calls for GPT-4o-mini, we conduct each experiment only once per dataset to ensure feasibility within a reasonable budget. This approach is common in agent recommendation studies \cite{Agent4Rec,AgentCF,RecAgent,recranker} and large-scale recommendation system research. Moreover, the larger number of users (1000) in our study enhances the reliability of the experimental results.

Note that we apply different LLM-UM methods to each dataset: Reflection for $\texttt{CDs}{50}$, and Summarization for $\texttt{CDs}{200}$ and $\texttt{Books}{480}$. The reason is that Reflection becomes inefficient as sequence length grows—a limitation also noted in the original AgentCF, and Summarization is more suitable for longer behavior sequence. 

\section{Backbone Methods}
\label{sec:app_backbone}
We provide a detailed description of the backbone methods used for validation.

\textbf{AgentCF \cite{AgentCF}} employs a reflective mechanism to model user personas. In the original framework, both the user profile and item profile are dynamically updated. In our implementation, the item profile is textually represented by concatenating the item's fields, while the user profile is initially set to "\texttt{Currently Unknown}" and is iteratively refined through continuous reflection. Furthermore, for the downstream recommendation ranking task in AgentCF, we replace the original LLM-based ranking with the EasyRec framework \cite{EasyRec}. EasyRec is the first large language embedding model specifically designed for recommendation. It aligns textual semantic spaces with collaborative behavioral signals, enabling recommendation tasks to rely solely on textual instructions (e.g., user preference descriptions and item profiles) while achieving performance comparable to traditional state-of-the-art models. Leveraging EasyRec for point-wise ranking is more experimentally efficient, accurate, and robust compared with LLMs.

\textbf{Agent4Rec \cite{Agent4Rec}} maintains an agent profile comprising two key components: social traits and unique tastes. In our implementation, we streamline the process by focusing solely on capturing diverse user interests through the construction of unique tastes, thus simplifying experimentation. To achieve this, we adopt the summarization method from the original work, which distills user preferences from their behavioral sequences. Additionally, we replace the original rating prediction task in the Agent4Rec framework with a ranking task.

\begin{figure*}[t]
  \centering
    \includegraphics[width=5.6 in]{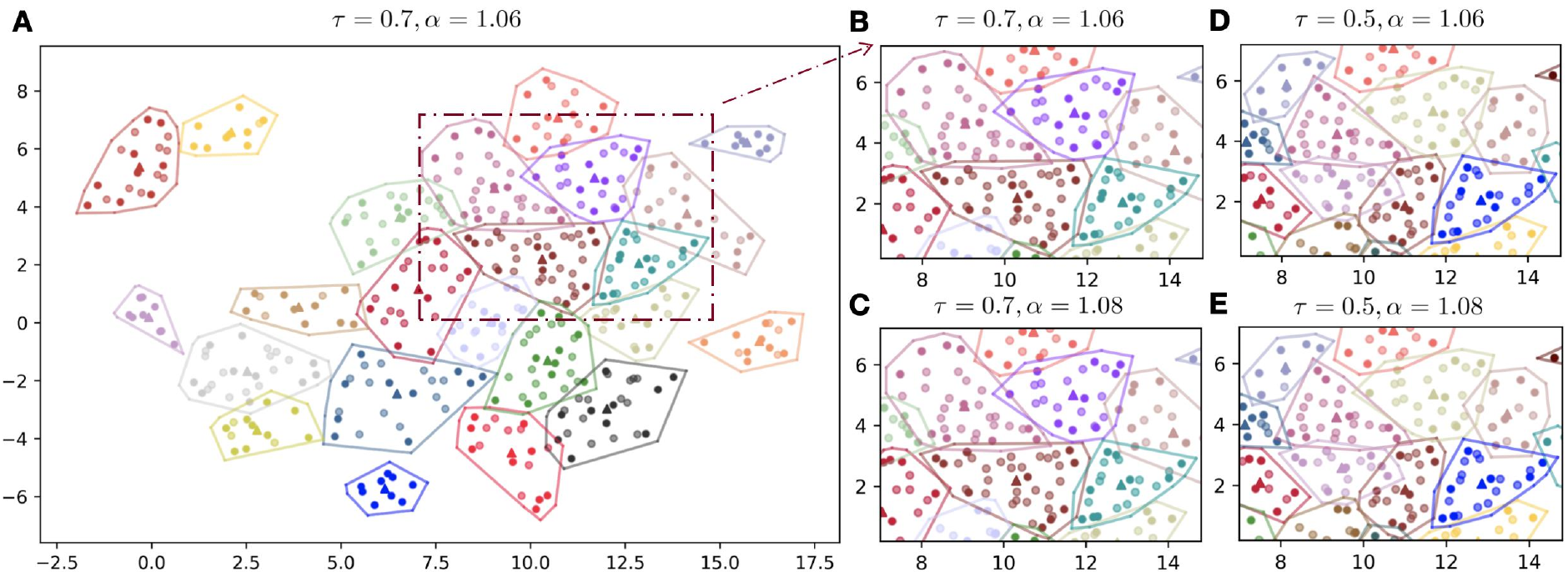}
   \caption{Sampling process for a user in $\texttt{Books}_{\texttt{480}}$ with a 50\% selection ratio. Points are color-coded and outlined. Non-transparent points signify data selected, whereas transparent points delineate behaviors not sampled. A offers a holistic perspective on the user's comprehensive behavior distribution, capturing the full extent of engagement patterns. B\~E presents parts of behaviors distributions and sampling process under varying configurations of hyper-parameters. Triangles denote the centroids of the clusters.}
    \label{fig: visualize}
\end{figure*}

\section{Hyper-parameter Analysis and Sampling Process Visualization}
\label{sec:app_hyper}
This section delves into the influence of the hyperparameters $\tau$ and $\alpha$ on the performance of PersonaX, as they play pivotal roles in shaping the hierarchical clustering and in-cluster behavior selection processes. Specifically, $\tau$ dictates the granularity of the hierarchical clustering. A larger $\tau$ value yields coarser clusters, encompassing a broader spectrum of behavioral samples with potentially greater divergence from the cluster centroid. In contrast, a smaller $\tau$ enforces a more stringent clustering criterion, resulting in finer-grained clusters characterized by higher intra-cluster homogeneity. On the other hand, $\alpha$ modulates the balance between prototypicality and diversity during the in-cluster behavior selection stage. A higher $\alpha$ amplifies the preference for selecting behavior samples further from the cluster centroid, thereby enhancing diversity within the cluster. Conversely, a lower $\alpha$ emphasizes prototypicality, favoring samples that closely align with the cluster centroid. Our empirical analysis, as illustrated in Figure~\ref{fig: hyper_parameter}, uncovers nuanced patterns in how these hyperparameters influence the model’s overall performance. 

\textbf{1. Performance at Low Ratios:} 
Across $\tau$ and $\alpha$ configurations, the performances at lower ratios (e.g., $0.1$, $0.3$) remain similar. This is because the selected samples at low ratios primarily originate near the cluster centroid, regardless of the diversity adjustment imposed by $\alpha$. Slightly superior performance of $\tau=0.5$ compared to $\tau=0.7$ at these ratios is attributed to the finer clustering granularity of $\tau=0.5$, which ensures that selected samples exhibit higher prototypicality.

\textbf{2. Performance at High Ratios ($0.5$–$0.9$):} 
At higher ratios, configurations with larger $\alpha$ values (e.g., $\alpha=1.06, 1.08$) outperform their smaller-$\alpha$ counterparts (e.g., $\alpha=1.01, 1.04$). This highlights the efficacy of the in-cluster selection strategy: after a core set of prototypical samples is chosen, incorporating more diverse samples significantly enhances performance. The inclusion of diversity helps capture broader behavioral patterns, leading to improved generalization.

\textbf{3. Trade-offs in Specific Settings:} 
A nuanced behavior is observed in the interaction between $\tau$ and $\alpha$. For $\tau=0.5$, $\alpha=1.08$ performs better than $\alpha=1.06$, suggesting that in scenarios where the cluster scope is relatively constrained, the diversity of samples becomes pivotal, necessitating a higher $\alpha$ to effectively prioritize and capture heterogeneous behaviors.  For $\tau=0.7$, $\alpha=1.06$ outperforms $\alpha=1.08$, as the broader cluster scope with $\alpha=1.08$ potentially overemphasizes highly diverse samples, leading to a slight degradation in overall performance. This interplay underscores the importance of balancing cluster granularity and diversity during sample selection.

\textbf{4. Parameter Robustness:}
Our framework demonstrates robust performance across a wide range of hyper-parameter settings. For instance, the worst best performance ($71.6$) achieved with $\tau=0.7, \alpha=1.04$ is only marginally lower than the best performance of the relevance baseline ($71.86$). This indicates that our method remains effective without being overly sensitive to hyper-parameter adjustments.

To provide an intuitive analysis of the sampling process, we conducted a visualization study, as illustrated in Figure~\ref{fig: visualize}. From Figure~\ref{fig: visualize}.A, it is evident that smaller clusters are preferentially allocated an adequate sampling quota compared to larger ones. This observation underscores the efficacy of the proposed Algorithm~\ref{alg:SamplingBudgetAllocation}, which strategically prioritizes smaller clusters to ensure sufficient sampling. By adopting this approach, the algorithm effectively preserves the user's diverse interests, including long-tail preferences, even under constrained sampling resources. The comparisons between Figure~\ref{fig: visualize}.B and Figure~\ref{fig: visualize}.C, as well as Figure~\ref{fig: visualize}.D and Figure~\ref{fig: visualize}.E, highlight the impact of $\alpha$. Specifically, smaller $\alpha$ values tend to focus the sample selection closer to the cluster centroids. Furthermore, the comparisons between Figure~\ref{fig: visualize}.B and Figure~\ref{fig: visualize}.D, and between Figure~\ref{fig: visualize}.C and Figure~\ref{fig: visualize}.E, demonstrate that a a more granular clustering can constrain Algorithm~\ref{alg:InClusterSelectionModified} from selecting samples that deviate excessively from the cluster centroids. This constraint mitigates potential performance degradation caused by overemphasis on unrelated samples.

The experimental findings and visualization analysis suggest that both $\tau$ and $\alpha$ require empirical tuning to identify optimal configurations. We recommended a balance between prototypicality and diversity, for example a larger $\alpha$ values combined with appropriately tuned small $\tau$.

\section{Details about In-Cluter Selection}
\label{sec:app_in_cluster}

In this section, we delve into the mechanisms governing sample selection by proposing a principled scoring system to evaluate the prototypicality and diversity of candidate samples. The scoring mechanism is derived from two complementary perspectives: prototypicality $$\frac{1}{1+d(\mathbf{e}_j, \mathbf{\mu}_i)}$$, which assesses how representative a sample is of its respective cluster, and diversity $$\frac{2}{a_i} \sum_{\substack{I_a, I_b \in c_i^* \\ a \neq b}} d(\mathbf{e}_a,\mathbf{e}_b)$$, which quantifies the extent to which the selected samples span a broader spectrum of the data distribution.

\subsection{Prototypicality and Diversity Scoring}
From the formulation below, 
{\footnotesize
$$
\begin{aligned}
    \max_{c_i^*} \Biggl( & w_p \cdot \sum_{I_j \in c_i^*} \frac{1}{1+d(\mathbf{e}_j, \mathbf{\mu}_i)} 
    + w_d \cdot \frac{2}{a_i} \sum_{\substack{I_a, I_b \in c_i^* \\ a \neq b}} d(\mathbf{e}_a,\mathbf{e}_b) \Biggr)
\end{aligned}
$$
}it is evident that the prototypicality score exhibits an inverse relationship with the distance between a sample and the center of its cluster. As a sample moves further from the cluster centroid, its prototypicality diminishes proportionally, reflecting its reduced ability to represent the typical characteristics of the cluster. The diversity score considers the pairwise distances between the candidate sample and the samples already selected. This ensures that the inclusion of a new sample enriches the diversity of the chosen subset by discouraging redundancy.

To compute the diversity score, we employ the scaling factor $2/a_i$. We don't choice of averaging scaling approach $1/[a_i(a_i-1)]$, which tends to normalize diversity growth. By adopting $2/a_i$, we deliberately amplify the influence of diversity as $a_i$ increases, thereby prioritizing the inclusion of diverse samples in scenarios where a cluster is allocated enough sampling budget. This design reflects an underlying intent: as $a_i$ grows, the system places greater emphasis on diversity to ensure comprehensive coverage of the data distribution. Conversely, when $a_i$ is small, prototypicality takes precedence, directing attention toward selecting samples that are most representative of their respective clusters.

\subsection{Design Rationale}
The decision to amplify diversity dynamically aligns with our broader goal of achieving a balanced and adaptive sample selection process. By coupling prototypicality with diversity in this manner, we address two critical challenges in data selection:

1. \textbf{Representative Sampling}: When the sample pool is sparse, selecting highly prototypical samples ensures that the chosen subset faithfully captures the core characteristics of the data clusters. This is particularly crucial in tasks where the representativeness of the selected data has a direct impact on model performance, such as user profiling or content recommendation.

2. \textbf{Comprehensive Coverage}: In cases where the candidate pool is dense, diversity becomes increasingly important to avoid redundancy and to capture the subtle variations within the data distribution. By amplifying diversity when $a_i$ is large, our scoring mechanism ensures that the selected subset spans the breadth of the distribution, enabling downstream models to generalize better across diverse scenarios.

\begin{figure}[t]
  \centering
    \includegraphics[width=2.5 in]{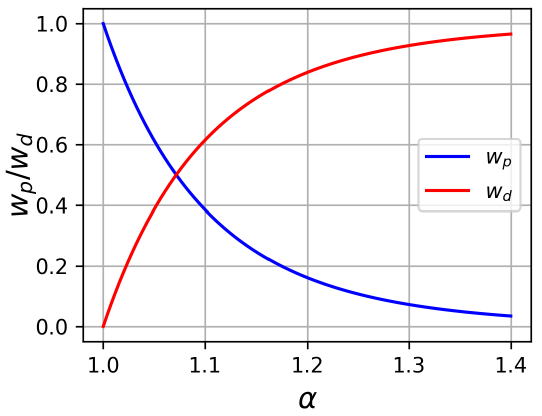}
   \caption{Trade off between $w_p$ and $w_d$ at different settings of $\alpha$.}
    \label{fig:w_p_w_d}
\end{figure}

\subsection{Broader Implications}
The proposed scoring framework introduces a novel perspective on balancing representativeness and diversity in data selection. By dynamically modulating the influence of diversity based on the local sample density, our approach strikes a principled balance between selecting typical and atypical samples. This adaptability is particularly valuable in data-centric applications, where sample selection directly affects the quality of downstream tasks, such as dataset pruning, user interest modeling, and few-shot learning.

\subsection{Visualization Explanation}
Figure~\ref{fig:w_p_w_d} shows the trade-off between $w_p$ and $w_d$ across different settings of $\alpha$. As observed in the figure, when $\alpha$ is small, $w_p$ dominates the sampling process, leading to the selection of samples near the cluster center. These samples are prototypical and reflect the representative thematic interests of the cluster. As $\alpha$ increases, $w_d$ becomes more prominent, and $w_p$ approaches 0, causing the sampling process to prioritize diverse samples in order to enhance generalization.

\begin{figure}[t]
  \centering
    \includegraphics[width=2.9 in]{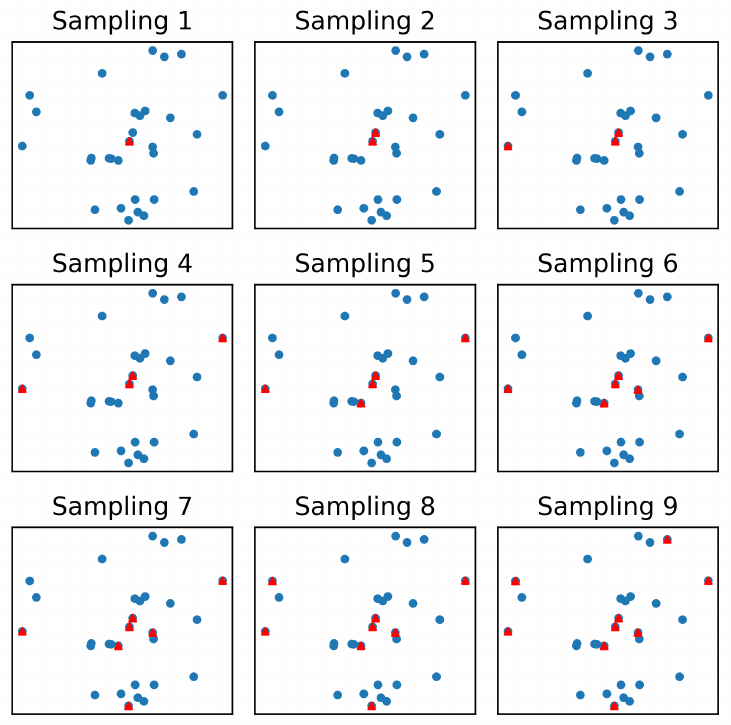}
   \caption{Dynamics of In-Cluster sample selection: We set $\tau=0.7$, with the samples distributed within the range of $[-2, 2]$.}
    \label{fig:in_cluster}
\end{figure}

Figure~\ref{fig:in_cluster} presents a dynamic visualization of the sampling process in Algorithm~\ref{alg:InClusterSelectionModified}. As illustrated, the algorithm iteratively selects samples by jointly optimizing for both prototypicality and diversity, thereby maximizing the combined gain. This approach stands in contrast to conventional data selection methods, which often exhibit a unimodal bias—either favoring simple, centrally clustered, and highly representative samples \cite{herding,ICARL,beyond} or prioritizing difficult, outlier samples with strong generalization potential \cite{diet,forgetting_score}.  

Empirical analysis of the hyperparameter $\alpha$, which governs the trade-off between prototypicality and diversity, reveals a practical range of $1.06$–$1.08$. Within this regime, PersonaX often firstly selects a minimal set of prototypical samples and then shifting its focus toward maximizing sample diversity. We believe this is because of the superior few-shot generalization capabilities of LLMs. These models inherently require fewer prototypical instances to capture core user interests, thereby shifting their emphasis toward diverse sample acquisition to further enhance generalization.

\begin{table}[t!]
\footnotesize
\centering
\caption{Intuitive comparison of user personas modeled by different methods. Specifically, (A) Relevance sampling (length 3), (B) Recent sampling (length 10), and (C) PersonaX (selection ratio 30\%).}
\label{tab:case_study}
\begin{tabular}{|c|p{3 in}|p{2.5 in}|}
\hline
\textbf{Method} & \textbf{User Persona} & \textbf{o3-mini's Reasoning Evaluation} \\ \hline
A & 
I prefer Rock and Progressive Rock music, particularly works by artists like Trey Anastasio. I enjoy albums that feature new songs and live performances, as they provide a dynamic listening experience. I appreciate behind-the-scenes content that adds depth to the music experience. I also enjoy Contemporary Folk music, especially works by artists like Steve Winwood, as they can offer a fresh take on musical compositions. I dislike R\&B and Soul music, particularly collections that focus on remixes of older tracks, as I find them less engaging. I prefer original compositions and fresh takes on music rather than revisiting past favorites. Additionally, I am not particularly interested in World Music, especially live performances that do not align with my preferred genres.
&
This description is detailed and rich in information, but it introduces multiple music genres which may dilute the focus on the user's core preferences. \\ \hline
B & 
I prefer Classic Rock and Arena Rock music, particularly albums that showcase strong melodies and engaging performances, such as 'Chickenfoot' by Chickenfoot. I enjoy music that has a fun, energetic vibe and often features collaborative efforts from talented musicians. I appreciate emotional ballads and songs that reflect depth and complexity in lyrics and composition. I dislike Blues Rock that focuses on traditional guitar work and may lack the innovative sounds I seek, as exemplified by 'Smokestacks, Broom Dusters \& Hoochie Coochie Men' by Micky Moody, which I find less appealing due to its more conventional approach.
&
This description focuses on a subset of rock music—Classic and Arena Rock—which contrasts with the broader rock and progressive preferences seen in the other descriptions. It is detailed but less aligned with the core focus compared to C. \\ \hline
C & 
I prefer rock and progressive music, particularly works by notable artists like Trey Anastasio. I enjoy albums that offer a collection of new songs, especially those that include additional content such as live performances and behind-the-scenes footage. I dislike pop and dance music, particularly generic albums that lack depth or a compelling narrative. I appreciate immersive listening experiences that connect me to the artist's journey and creative process.
&
This description is the most concise and focused, effectively capturing the user's core interests—new material, live performances, and behind-the-scenes insights—without extraneous details, making it the highest quality among the three. \\ \hline
\end{tabular}
\end{table}

\begin{table}[t!]
\footnotesize
\centering
\caption{Quantitative Evaluation of User Persona Modeling Methods}
\label{tab:quantitative_results}
\begin{tabular}{@{}lccccccc@{}}
\toprule
\textbf{Method} & \textbf{NDCG@1} & \textbf{NDCG@5} & \textbf{NDCG@10} & \textbf{Hit@1} & \textbf{Hit@5} & \textbf{Hit@10} & \textbf{MRR} \\ \midrule
A & 0.00 & 0.42 & 0.54 & 0.00 & 0.67 & 1.00 & 0.39 \\
B & 0.00 & 0.54 & 0.54 & 0.00 & 1.00 & 1.00 & 0.39 \\
C & 0.33 & 0.71 & 0.71 & 0.33 & 1.00 & 1.00 & 0.61 \\ \bottomrule
\end{tabular}
\label{tab:case_study_metric}
\end{table}

\section{Theoretical Analysis for In-cluster Selection}
\label{sec:theo_analysis}
Algorithm~\ref{alg:InClusterSelectionModified}
(\emph{P}rototypicality–\emph{D}iversity
balanced \emph{S}ub-\emph{B}ehaviour \emph{S}equence) selection
greedily builds an SBS~$S_t\subseteq V_i$ of cardinality~$t$
by adding at each iteration the element with the largest marginal gain
with respect to the mixed objective~\eqref{eq:objective}.
We establish three properties: monotonicity, finite termination,
and a data–dependent $\,94.79\%$ approximation guarantee.

\subsection{Objective Function Properties}

The following is the in-cluster SBS selection objective function for a fixed cluster~$c_i^{\ast}$:
\begin{equation}
\label{eq:objective}
\max_{c_i^{\ast}}
\; f(c_i^{\ast})
\;=\;
w_p \,\underbrace{\sum_{I_j \in c_i^{\ast}}
\frac{1}{1+d\!\left(\mathbf{e}_j,\boldsymbol{\mu}_i\right)}}_{\displaystyle f_p(c_i^{\ast})}
\;+\;
w_d \,\underbrace{\frac{2}{a_i}
\sum_{\substack{I_a,I_b \in c_i^{\ast}\\a\neq b}}
d\!\left(\mathbf{e}_a,\mathbf{e}_b\right)}_{\displaystyle f_d(c_i^{\ast})},
\end{equation}
where $d(\cdot,\cdot)$ denotes a metric in the embedding space,
$\boldsymbol{\mu}_i$ is the centroid of cluster~$i$, $a_i$ is sampling size,
and $w_p,w_d\!\ge\!0$ are fixed weights.

\vspace{0.5em}
\begin{definition}
\label{def:submodular}
Let $V$ be a finite ground set.  
A set function $f\colon 2^{V}\!\to\!\mathbb{R}$ is \emph{submodular} if for all
$A\subseteq B\subseteq V$ and for every $e\in V\setminus B$,
\begin{align}
f(A\cup\{e\})-f(A)\;\;\ge\;\;f(B\cup\{e\})-f(B).
\end{align}
A set function is \emph{supermodular} when the inequality is reversed,
and \emph{modular} when equality always holds.
\end{definition}

\begin{lemma}
\label{lem:prototypicality_modular}
The prototypicality component $f_p$ in~\eqref{eq:objective} is modular
and hence submodular.
\end{lemma}

\begin{proof}
For any subset $S\subseteq V$,
\[
f_p(S)=\sum_{I_j\in S}\frac{1}{1+d\!\left(\mathbf{e}_j,\boldsymbol{\mu}_i\right)}
=\sum_{I_j\in S}g(I_j),
\]
where $g(I_j)$ depends solely on the singleton $I_j$.  
Thus the marginal gain of adding $e$ is always $g(e)$,
independent of~$S$, satisfying the equality condition in
Definition~\ref{def:submodular}.
\end{proof}

\begin{lemma}
\label{lem:diversity_supermodular}
The diversity component $f_d$ in~\eqref{eq:objective} is supermodular.
\end{lemma}

\begin{proof}
Let $A\subseteq B\subseteq V$ and $e\in V\setminus B$.  
Denote $m_A(e)=f_d(A\cup\{e\})-f_d(A)$ and
$m_B(e)=f_d(B\cup\{e\})-f_d(B)$.  By direct expansion,
\[
m_A(e)=\frac{2}{a_i}\sum_{a\in A}d\!\left(\mathbf{e},\mathbf{e}_a\right),
\qquad
m_B(e)=\frac{2}{a_i}\sum_{b\in B}d\!\left(\mathbf{e},\mathbf{e}_b\right).
\]
Because $A\subseteq B$, every term in the sum for $m_A(e)$
appears in $m_B(e)$ (and $m_B(e)$ contains additional non-negative
terms due to metric non-negativity).  Hence $m_A(e)\le m_B(e)$,
which is exactly the reverse inequality of
Definition~\ref{def:submodular}, establishing supermodularity.
\end{proof}

\begin{proposition}
\label{prop:overall_mixed}
The full objective~$f=w_p f_p + w_d f_d$ is the weighted sum of a modular
(submodular) function and a supermodular function.  
Consequently $f$ itself is \emph{neither} submodular nor supermodular
unless $w_d=0$ or $w_p=0$, respectively.
\end{proposition}

\begin{proof}
Immediate from Lemmas~\ref{lem:prototypicality_modular}
and~\ref{lem:diversity_supermodular} and the linearity of set
functions.
\end{proof}

\paragraph{Monotonicity.}
Every candidate element has non-negative marginal gain,
the greedy rule---selecting at iteration $t$ the element with the
largest marginal improvement---yields a sequence of objective values
$
f(S_{t+1}) \;>\; f(S_t)
$
until the prescribed cardinality~$a_i$ is reached.

\paragraph{Finite Termination.}
The procedure performs exactly $a_i$ iterations, inserting one
element per step; therefore it terminates after a finite number of
steps.

\subsection{Performance of the Greedy Algorithm}
\label{sec:greedy_analysis}

Bian et al.~\cite{greed_is_good} study the maximisation of monotone
\emph{BP} functions—sums of a monotone submodular component~$f$ and a
monotone supermodular component~$g$—under a cardinality constraint.
Let\footnote{We denote
$\displaystyle f(v)=f(\{v\})$ and
$f(v \mid S)=f(S\cup\{v\})-f(S)$ for brevity.}
\begin{align}
\kappa_g
&\;=\;
1-\min_{v\in V}
\frac{g(v)}
     {g\!\bigl(v \mid V\setminus\{v\}\bigr)},
&
\kappa_f
&\;=\;
1-\min_{v\in V}
\frac{f\!\bigl(v \mid V\setminus\{v\}\bigr)}
     {f(v)}.
\label{eq:curvatures}
\end{align}
The quantities $\kappa_g\in[0,1]$ and $\kappa_f\in[0,1]$ are termed
the \emph{curvatures} of~$g$ and~$f$, respectively, and capture how
far each component deviates from modularity.  For monotone BP
functions the simple greedy algorithm that, at every step, adds the
element of highest marginal gain enjoys the worst-case guarantee.
\begin{equation}
\label{eq:bp_bound}
\frac{F_{\text{greedy}}}{F^\star}
\;\ge\;
\frac{1}{\kappa_f}
\Bigl(1-\exp\!\bigl(-\kappa_f(1-\kappa_g)\bigr)\Bigr),
\end{equation}
where $F^\star$ is the optimal objective value.

\paragraph{Empirical curvatures.}
Using the settings $\tau=0.7$ and $\alpha=1.06$,
we compute the pointwise ratios
$
r_v^g
=
\frac{g(v)}{g(v \mid V\setminus\{v\})}
$
and
$
r_v
=
\frac{f(v \mid V\setminus\{v\})}{f(v)}
$
for the ten most representative items.
Table~\ref{tab:ratio_modular} reports the results (smaller
ratios---\emph{bold} in the table---produce larger curvatures).

\begin{table}[t]
\centering
\caption{Pointwise ratios for the supermodular ($g$) and submodular ($f$)
components; lower ratios imply higher curvature.
Empirically SBS selection only has $a_i\le5$, we \textbf{highlight} the values with attaining the minimum.}
\label{tab:ratio_modular}
\begin{tabular}{ccc}
\toprule
Point &
$\displaystyle r_v^g \!=\! \frac{g(v)}{g(v \mid V\setminus\{v\})}$ &
$\displaystyle r_v \!=\! \frac{f(v \mid V\setminus\{v\})}{f(v)}$\\
\midrule
1 & 0.0765 & 0.1648 \\
2 & 0.0822 & 0.2075 \\
3 & 0.0666 & 0.1509 \\
4 & \textbf{0.0493} & \textbf{0.1159} \\
5 & 0.0884 & 0.2137 \\
\hline
6 & 0.0599 & 0.1374 \\
7 & 0.0847 & 0.2028 \\
8 & 0.0301 & 0.0915 \\
9 & 0.0851 & 0.2056 \\
10 & 0.0194 & 0.0696 \\
\bottomrule
\end{tabular}
\end{table}

Taking the minima over all points gives
$\kappa_g\!\le\!0.9806$ and $\kappa_f\!\le\!0.9304$.
Substituting these into~\eqref{eq:bp_bound} yields a
\emph{guaranteed} approximation factor of $94.79\%$ for our
in-cluster greedy selector (Algorithm~\ref{alg:InClusterSelectionModified}).

\paragraph{Conclusion.}
Despite the lack of submodularity,
the curvature-aware guarantee shows that the proposed greedy
in-cluster selection is near-optimal in practice.

\section{Case Study} \label{app:app_case_study}
In this section, we present a case study comparing user personas modeled using Relevance, Recent, and PersonaX methods, with the backbone LLM-UM approach fixed as Reflection. The dataset used is $\texttt{CDs}_{50}$, with the User ID \texttt{A2NQUGGYM0DBM1}. The results are summarized in Table~\ref{tab:case_study}. We evaluate their quality by OpenAI's o3-mini, using its reasoning capabilities in an LLM-As-Judge framework. The evaluation indicated that Model C had the highest modeling quality \footnote{Repeated inquiries occasionally resulted in A being rated higher, with the justification that A offered a more comprehensive view. However, this comprehensiveness came at the cost of interest modeling that was more diffuse and less precise.}. The explanation provided was that C demonstrated superior descriptive quality, capturing the user’s core preferences for rock and progressive music with concise and precise language. It also emphasized the user’s interest in new releases, live performances, and behind-the-scenes content, while avoiding extraneous information misaligned with primary interests. In contrast, Model A, while rich in information, introduced a broader range of music styles that diluted focus, and Model B predominantly emphasized an alternative style of rock, leading to inconsistencies with the other descriptions.

We conducted three rounds of quantitative evaluations on the ranking task, each comprising one positive item alongside nigh negative items. As shown in Table~\ref{tab:case_study_metric}, Method C achieved the highest performance, followed by Method B, while Method A exhibited the poorest performance.

\section{Prompt Templates}
\label{sec:app_prompts}
We present the prompt templates used in AgentCF, as shown in Figure~\ref{fig:agentcf_forward} and Figure~\ref{fig:agentcf_backward}, and those employed in Agent4Rec, depicted in Figure~\ref{fig:agent4rec_summ}.

\begin{lrbox}{\InterviewCase}
\begin{tcolorbox}[colback=background_u,colframe=frame_u,boxrule=0.5pt,title={\textbf{Prompt Template for Forward Inference Process of AgentCF}},width=\textwidth]
\textbf{Task:}
We provide a user's personal profile in [User Profile], which includes the user's preferences, dislikes, and other relevant information. You need play the role of the user. And we also provide two candidate items, A and B, with their features in [Item Feature].  You need to choice between the two item candidates based on your profile and the features of the items. Furthermore, you must articulate why you’ve chosen that particular item while rejecting the other.\\
\textbf{User Profile:} 
\{profile\}\\
\textbf{Item Feature:} 
Item A: \{item a\} Item B: \{item b\} \\
\textbf{Steps to Follow:} 

1. Extract your preferences and dislikes from your self-introduction.

2. Evaluate the two candidate in light of your preferences and dislikes. Make your choice by considering the correlation between your preferences/dislikes and the features of the candidates.

3. Explain why you made such choices, from the perspective of the relationship between your preferences/dislikes and the features of these candidate items.\\
\textbf{Important Notes:} 

1. Your output should strictly be in the following format:
    Chosen Item: {{Item A or Item B}}
    Explanation: {{Your detailed rationale behind your choice and reasons for rejecting the other item.}}
    
2. When identifying user's likes and dislikes, do not fabricate them! If your [User Profile] doesn’t specify any relevant preferences or dislikes, use common knowledge to inform your decision.

3. You **must** choose one of these two candidates, and **cannot** choose both.

4. Your explanation needs to be comprehensive and specific. Your reasoning should delve into the finer attributes of the items.

5. Base your explanation on facts. For instance, if your self-introduction doesn’t reveal any specific preferences or dislikes, justify your decision using available or common knowledge.

6. Please ignore the effect of Item position and length, they do not affect your decision.\\
\textbf{Response Example:} 
\textit{Chosen Item: Item A
Explanation: I chose Item A because...}\\

\end{tcolorbox}
\end{lrbox}

\begin{figure*}[h]
    \centering
    \usebox{\InterviewCase}
    \vspace{-15pt}
    \caption{Prompt template for the forward process of AgentCF to predict one user potentially liked item between a positive one and a negarive one.}
    \label{fig:agentcf_forward}
    \vspace{-15pt}
\end{figure*}

\begin{lrbox}{\InterviewCase}
\begin{tcolorbox}[colback=background_u,colframe=frame_u,boxrule=0.5pt,title={\textbf{Prompt Template for Backward Reflection Process of AgentCF}},width=\textwidth]
\textbf{Background:}
We provide a user's personal profile in [User Profile], which includes the user's preferences, dislikes, and other relevant information. You need play the role of the user. Recently, you considered choosing one more prefered Item from two candidates. The features of these two candidate are provided in [Item Feature]. And your choice and explanation is in [Choice and Explanation], which reveals your previous judgment for these two candidates.\\
\textbf{User Profile:} 
\{profile\}\\
\textbf{Item Feature:} 
Item A: \{item a\} Item B: \{item b\} \\
\textbf{Choice and Explanation:} 
\{response\}\\
\textbf{Task:}
However, The user in the real world actually prefer to choose Item B, and reject the Item A that you initially chose. This indicates that you made an incorrect choice, the [Choice and Explanation] was mistaken. Therefore, you need to reflect and update [User Profile]. \\
\textbf{Steps to Follow:} 

1. Analyze the misconceptions in your previous [Choice and Explanation] about your preferences and dislikes, as recorded in your explanation, and correct these mistakes.  

2. Explore your new preferences based on the Item B you really enjoy, and determine your dislikes based on the Item a you truly don’t enjoy.  

3. Summarize your past preferences and dislikes from your previous [User Profile]. Combine your newfound preferences and dislikes with your past ones. Filter and remove any conflicting or repetitive parts in your past [User Profile] that contradict your current preferences and dislikes.  

4. Generate a update profile use the following format: 

My updated profile: \{Please write your updated profile here\}\\
\textbf{Important Notes:} 

1. Keep your updated profile under 180 words.  

2. Any overall assessments or summarization in your profile are forbidden.  

3. Your updated profile should only describe the features of items you prefer or dislike, without mentioning your wrong choice or your thinking process in updating your profile.  

4. Your profile should be specific and personalized. Any preferences and dislikes that cannot distinguish you from others are not worth recording.\\
\textbf{Response Example:} 
\textit{My updated profile: I ...}\\

\end{tcolorbox}
\end{lrbox}

\begin{figure*}[t]
    \centering
    \usebox{\InterviewCase}
    \vspace{-15pt}
    \caption{Prompt template for the backward process of AgentCF to apply the reflect mechanism for updating user profile.}
    \label{fig:agentcf_backward}
    \vspace{-15pt}
\end{figure*}

\begin{lrbox}{\InterviewCase}
\begin{tcolorbox}[colback=background_u,colframe=frame_u,boxrule=0.5pt,title={\textbf{Prompt Template for Summarization Process of Agent4Rec}},width=\textwidth]
\textbf{Task:}
We provide a user's personal profile in [User Profile], which includes the user's preferences and other relevant information. Additionally, we provide a sequence of liked items in [Sequence Item Profile] that the user has interacted with. Your task is to analyze these items in the context of the user's existing profile and produce an updated profile that reflects any new preferences, or insights inferred from the user's interactions with these items.\\
\textbf{User Profile:} 
\{profile\}\\
\textbf{Sequence Item Profile:} 
\{sequence item profile\}\\
\textbf{Steps to Follow:} 

1. Carefully review the user's existing profile to understand their stated preferences and dislikes.

2. Analyze the features of the items in the provided sequence, noting any common themes, attributes, or patterns.

3. Identify any new preferences that can be inferred from the user's interactions with these items.

4. Summarize and update the user's profile by incorporating the new insights, adding new preferences or dislikes, and highlighting any changes or developments in the user's tastes.
Important Notes

5. Your output should strictly be in the following format:
Summarization: \{Your updated profile.\}

6. Do not contradict the user's existing preferences unless there is clear evidence from the sequence items that their tastes have changed.

7. Base your summary on facts and logical inferences drawn from the items in the sequence.

8. Be comprehensive and specific in your summarization, focusing on the finer attributes and features of the items that relate to the user's preferences.

9. Avoid fabricating any information not supported by the user's profile or the sequence items.\\
\textbf{Response Example:}
Summarization: You've developed interest in .... \\
\end{tcolorbox}
\end{lrbox}

\begin{figure*}[t!]
    \centering
    \usebox{\InterviewCase}
    \vspace{-15pt}
    \caption{Prompt template of Agent4Rec to apply the summarization mechanism for distilling user profile.}
    \label{fig:agent4rec_summ}
    \vspace{-15pt}
\end{figure*}

\end{document}